\documentclass{aastex63}
\usepackage{color}
\usepackage{bm}
\usepackage{cases}
\usepackage{multirow}
\usepackage{lineno}

\received{}
\revised{}
\accepted{}
\submitjournal{}
\shorttitle{}
\shortauthors{}
\begin{document}

    \title{The improved Amati correlations from Gaussian copula}

    \author{Yang Liu}
      \affiliation{Department of Physics and Synergistic Innovation Center for Quantum Effects and Applications, Hunan Normal University, Changsha, Hunan 410081, China}
       
   \author{Fuyong Chen}
    \affiliation{Department of Physics and Synergistic Innovation Center for Quantum Effects and Applications, Hunan Normal University, Changsha, Hunan 410081, China}
    \author{Nan Liang}
    \affiliation{Key Laboratory of Information and Computing Science Guizhou Province, Guizhou Normal University, Guiyang, Guizhou 550025, China}
   \affiliation{Yunnan Observatory, Chinese Academy of Sciences, Kunming, Yunnan 650011, China}

    \author{Zunli Yuan}
    \affiliation{Department of Physics and Synergistic Innovation Center for Quantum Effects and Applications, Hunan Normal University, Changsha, Hunan 410081, China}

    \author{Hongwei  Yu}
    \affiliation{Department of Physics and Synergistic Innovation Center for Quantum Effects and Applications, Hunan Normal University, Changsha, Hunan 410081, China}

    \author{Puxun Wu}
    \affiliation{Department of Physics and Synergistic Innovation Center for Quantum Effects and Applications, Hunan Normal University, Changsha, Hunan 410081, China}
 
%

\begin{abstract}
In this paper, we obtain two improved Amati correlations of the Gamma-Ray burst (GRB) data via a powerful statistical tool called copula. After calibrating, with the low-redshift GRB data, the improved Amati correlations based on a fiducial $\Lambda$CDM model with $\Omega_\mathrm{m0}=0.3$ and $H_0=70~\mathrm{km~s^{-1}Mpc^{-1}}$,  and extrapolating the results to the high-redshift GRB data,  we obtain the Hubble diagram of GRB data points.  Applying these GRB data to constrain the $\Lambda$CDM model, we find that the improved Amati correlation from copula  can give a result well consistent with $\Omega_\mathrm{m0}=0.3$, while the standard Amati and extended Amati correlations  do not. This results suggest that when the improved Amati correlation from copula is used in the low-redshift calibration method, the GRB data can be regarded as a viable cosmological explorer. However, the Bayesian information criterion indicates that the standard Amati correlation remains to be favored mildly since it has the least model parameters. Furthermore,  once the simultaneous fitting method rather than the low-redshift calibration one is used,    there is no apparent evidence that the improved Amati correlation is better than the standard one. Thus, more works need to be done in the future in order to compare different Amati correlations.

\end{abstract}

\section{Introduction}
The accelerating cosmic expansion was first discovered  by two independent teams in 1998 \citep{Riess1998,Perlmutter1999} from observation of Type Ia supernova (SNIa).
This discovery was further confirmed by many other observations, including the cosmic microwave background radiation (CMB)~\citep{CMB1,CMB2} and the baryon acoustic oscillation(BAO)~\citep{BAO}. To explain this observational phenomenon, a mysterious dark energy  has been proposed to exist in our universe. The nature of dark energy can be characterized by its equation of state (EoS) parameter.
The simplest candidate of dark energy is the cosmological constant $\Lambda$, whose EoS parameter equals to $-1$, and the  cosmological constant plus cold dark matter ($\Lambda$CDM) model is consistent with observational data very well. 
Based on the $\Lambda$CDM model,  a tight constraint on the Hubble constant $H_0$ can be given with very high-redshift  CMB data \citep{Planck},  which has a more than $4\sigma$  deviation from the value  of $H_0$ obtained directly from the very low-redshift SNIa data~\citep{Riess2021,Riess2018,Riess2018M}. Some other low-redshift observational data such as the Hubble parameter measurements $H(z)$, BAO, and the strong gravitational lenses, have been used to explore the Hubble constant, and given a lower $H_0$ than the one from the SNIa with large error bars which are reasonably consistent with the value of Planck 2018~\citep{Wu, Chen2017,Abbott2018,Birrer2020,Cao2021a, Lin2021,Khetan2021, Efstathiou2020,Freedman2021}. The $H_0$ tension seems to suggest that the assumed $\Lambda$CDM model used to determine the Hubble constant may be inconsistent with our present universe or  there are potentially unknown systematic errors in observational data. 
To precisely nail down the property of dark energy and  the possible origin of $H_0$ tension, intermediate-redshift observational data are necessary, since the SNIa and BAO only contain the redshift ranges of $z\lesssim 2$ and the CMB is at $z\sim 1100$.

Gamma-Ray bursts (GRBs) are the most violent and intriguing explosions in the universe, whose power is dominant in the (sub-)$\mathrm{MeV}$ gamma-ray range~\citep{Klebesadel1973}. The highest isotropic energy of GRBs can be up to $10^{54}~\mathrm{erg}$ and the detected redshift of GRBs reaches to $z\sim9.4$. Thus, GRBs have the potential to be a farther tracker than SNIa to explore the cosmic evolution. Many empirical luminosity correlations which are relationships between parameters of the light curves and/or spectra with the GRB luminosity or energy have been proposed to standardize GRB samples, such as the time lag-isotropic peak luminosity correlation ($\tau_{lag}-L_{iso}$) \citep{Norris2000}, the time variability-isotropic peak luminosity correlation ($V-L_{iso}$) \citep{Fenimore2000}, the Amati correlation (see the Appendix \ref{Amati} for the details) which connects the spectral peak energy in GRB cosmological rest-frame and the the isotropic equivalent radiated energy ($E_p-E_{iso}$)~\citep{Amati2002}, the correlation between the spectral peak energy and the isotropic peak luminosity ($E_p-L_{iso}$) \citep{Yonetoku2004}),  the correlation between the spectral peak energy and the collimated-corrected energy ($E_p-E_\gamma$) \citep{Ghirlanda2004}, and so on. Moreover, the issue of whether these  luminosity correlations are redshift dependent is investigated in \citep{Basilakos2008,Demianski2017,Wang2011, Li2007, Lin2015, Lin2016,Khadka2021}, and .  For an example, \citet{Demianski2017} introduced a power-law function to describe the evolutionary function of  the Amati correlation with redshift, and \citet{Wang2017} selected two redshift dependent formulas to parameterize the coefficients in the Amati correlation, which is called the extended Amati correlation in this paper (the details can be found in the Appendix \ref{Appendix_exAmati}).  
Since the luminosity $L_{iso}$, the isotropic energy $E_{iso}$ and the collimated-corrected energy $E_\gamma$ in GRBs are cosmology-dependent, the Hubble diagram can be obtained from standardized GRB samples.

To use the GRBs as a cosmological probe, we need to calibrate GRB correlations. Let us point out here that the cosmology-dependent calibration method suffers the so-called \textit{circularity problem}~\citep{Ghirlanda2006,Wang2015}. To avoid this problem, a \textit{low-redshift method}~\citep{Liang2008, Kodama2008, Wei2009, Liang2010}  to calibrate the GRB correlations in a cosmology-independent way was proposed, which uses  other distance probes such as SNIa to calibrate the GRBs at
low-redshifts and then extrapolates the results to high-redshifts to constrain the cosmological parameters. 
For this method, since the intrinsic dispersion of SNIa data is very much smaller than the GRB intrinsic dispersion, the SNIa data will dominate over the GRB data in a joint analysis of them on the cosmological constraints and so the resulting cosmological constraints are effectively from SNIa data.
On the other hand, the \textit{simultaneous fitting} or \textit{global fitting}~\citep{Ghirlanda2004b,Li2008} limits the coefficients of the luminosity correlations and the parameters of cosmological models simultaneously from the observational GRB data. 
In \citep{Khadka2020, Khadka2021}, by simultaneously fitting cosmological and GRB Amati correlation parameters and using a number of different cosmological models, it has been found that the Amati correlation parameter values are independent of the cosmological model, which seems to mean that there is no circularity problem and that these GRB data sets are standardizable within the error bars. Up to now,  the GRB data has been used widely to investigate different cosmological models~\citep[see recent Refs:][]{Amati2019,Demianski2021,Cao2021b,Cao2022,Hu2021,Wang2021,Luongo2021}.

In this work we aim to improve the Amati correlation which has been  widely used in GRB cosmology.
We introduce a powerful statistical tool called \textit{copula}, which is a specialized tool developed in modern statistics to describe the complicated dependence structures between multivariate random variables. Copula has been widely used in various areas such as mathematical finance and hydrology in the past few decades.
In recent years, it has gradually been recognized by the astronomical community as a very useful tool to analyze data.
For example, \citet{Yuan2018} successfully determined the luminosity function of the radio cores in active galactic nuclei via copula, which is  difficult to do if the traditional method is used, 
\citet{Koen2009}  studied, using copula,  the correlation between the GRB peak energy and the associated supernova peak brightness, \citet{Benabed2009} proposed a new approximation for the low multipole likelihood of the CMB temperature, and \citet{Jiang2009} constructed a period-mass function for extrasolar planets.
In addition, the copula likelihood function was  constructed for the convergence power spectrum from the weak lensing surveys by~\citet{Sato2010,Sato2011}. Modeling  bivariate astronomical data with copula instead of the conventional gaussian mixture method was  proposed by~\citet{Koen2017}. The copula function is also useful  in the study of galaxy luminosity functions and the large scale structure fields of matter density~\citep{Takeuchi2010,Takeuchi2013,Takeuchi2020,Scherrer2010,Qin2020}.
Based on the 3-dimensional Gaussian copula, we propose,  in this work,  two improved Amati correlations  and compare them with the standard Amati correlation and the extended Amati correlation by using the latest GRB samples~\citep{Khadka2021}.

The frame of this paper is arranged as follows: Section \ref{Sec_copula} studies the issue of how to construct a 3-dimensional probability density function (PDF) through a Gaussian copula. Two improved Amati correlations are obtained in Section \ref{Sec_imp_Amati}. In Section \ref{Sec_result}, a comparison between the improved Amati correlations and the (extended) Amati correlation is made by using both low-redshift calibration and simultaneous fitting methods. The conclusions are given in Section \ref{Sec_conclusion}.

\section{The copula}\label{Sec_copula}

 Briefly speaking, copulas are functions that join or ``couple'' multivariate distribution functions to their one-dimensional marginal distribution functions ~\citep{Nelson2006}.
Let $x$, $y$ and $z$ be three random variables, we use  $F(x)$, $G(y)$ and $W(z)$ to express their marginal cumulative distribution functions (CDFs) respectively, and $H$ as the joint distribution function of the three variables. According to Sklar's theorem, if $F$, $G$ and $W$ are continuous, there exists a unique copula $C$ such that
\begin{equation}
\label{Copula1}
H(x,y,z;\bm{\theta})=C\left (F(x),G(y),W(z);\bm{\theta} \right),
\end{equation}
where $\bm{\theta}$ denotes the parameters of the copula function $C$. Let $u\equiv F(x),~v\equiv G(y)$, and $q\equiv W(z)$, the joint probability density function (PDF) $h(x,y,z;\bm{\theta})$ can be obtained by
\begin{eqnarray}\label{pdf_h}
h(x,y,z;\bm{\theta})&=&\frac{\partial^3 H(x,y,z;\bm{\theta})}{\partial x\partial y\partial z}\nonumber\\
&=&\frac{\partial^3 C(u,v,q;\bm{\theta})}{\partial u\partial v\partial q}\frac{\partial u}{\partial x}\frac{\partial v}{\partial y}\frac{\partial q}{\partial z}\nonumber\\
&=&c[u,v,q;\bm{\theta}]f(x)g(y)w(z),
\end{eqnarray}
where $f(x)$, $g(y)$ and $w(z)$ are the marginal PDFs of $x$, $y$ and $z$ respectively,  and $c(u,v,q;\bm{\theta})$ is the density function of $C$.

One obvious advantage of Eq.~(\ref{Copula1}) is that by using copulas one can model the dependence structure and the marginal distributions separately.
All the information on the dependence between the three variables is carried by the copula \citep[see, e.g.,][]{Yuan2018}.
The next issue  is to find a 3-dimension  optimal copula function and  estimate its parameters to describe the observed data in GRBs. The bivariate copulas are abundant and thus the 3-dimensional copula functions are abundant too since most of the 2-dimensional copula can be extended easily to the 3-dimensional case.
Here we use the 3-dimensional Gaussian copula function with linear correlation  coefficient $\bm{\theta}=\{\rho_1,\rho_2,\rho_3\}$ to model our data, i.e.,
\begin{eqnarray}
C(u,v,q;\bm{\theta})=\Psi_3\left[\Psi_1^{-1}(u),\Psi_1^{-1}(v),\Psi_1^{-1}(q);\bm{\theta}\right],
\end{eqnarray}
where $\Psi_3$ and $\Psi_1$ are the standard 3-dimensional Gaussian CDF and 1-dimension Gaussian CDF respectively, and  $\Psi_1^{-1}$ denotes the inverse of $\Psi$.
 The density function of the Gaussian copula can be obtained from
\begin{eqnarray}\label{gauss_c}
c(u,v,q;\bm{\theta})&=&\frac{\partial^3 \Psi_3\left[\Psi_1^{-1}(u),\Psi_1^{-1}(v),\Psi_1^{-1}(q);\bm{\theta}\right]}{\partial u\partial v\partial q}\nonumber\\
&=&\frac{1}{\sqrt{\mathrm{det}\mathbf{\Sigma}}}\exp\left\{-\frac{1}{2} \left[\mathbf{\Psi^{-1}}^{T}\left(\mathbf{\Sigma}^{-1}-\mathbf{I}\right)\mathbf{\Psi^{-1}}\right]\right\},
\end{eqnarray}
where $\mathbf{\Psi^{-1}}\equiv\left[\Psi^{-1}(u),\Psi^{-1}(v),\Psi^{-1}(q)\right]^T$ , $\mathbf{I}$ stands for the identity matrix, and $\mathbf{\Sigma}^{-1}$ denotes the inverse of the covariance matrix $\mathbf{\Sigma}$, which reads
\begin{eqnarray}
\mathbf{\Sigma}\equiv
\left(\begin{array}{ccccc}
1~ & \rho_1~ & \rho_2 \\
\rho_1~ & 1~ & \rho_3 \\
\rho_2~ & \rho_3~ & 1
\end{array}\right).
\end{eqnarray}

The conditional PDF of $y$ denotes the probability of variable $y$  when $x$ and $z$ are fixed, which can be expressed as:
\begin{eqnarray}\label{cond_h}
g_y(y|x,z;\bm{\theta})&\equiv&\frac{h(x,y,z;\bm{\theta})}{h_{xz}(x,z;\rho_2)}
=\frac{c[u,v,q;\bm{\theta}]f(x)g(y)w(z)}{c[u,q;\rho_2]f(x)w(z)} =\frac{c[u,v,q;\bm{\theta}]}{c[u,q;\rho_2]}g(y),
\end{eqnarray}
where   $h_{xz}(x,z;\rho_2)$  is constructed from a 2-dimensional Gaussian copula with the correlation coefficient being $\rho_2$. If variable $y$ obeys a Gaussian distribution with the standard deviation being $\sigma_y$, then $g_y$ can be expressed as
\begin{eqnarray}\label{cond_y}
g_y(y|x,z;\bm{\theta})&=&\frac{1}{\sqrt{2\pi\sigma_{y|x,z}^2}}\exp\left[-\frac{1}{2}S(x,y,z;\bm{\theta})\right] ,
\end{eqnarray}
where
\begin{eqnarray}
\sigma_{y|x,z}^2&\equiv &\frac{\sigma_y^2\left(1-\rho_1^2-\rho_2^2-\rho_3^2+2\rho_1\rho_2\rho_3\right)}{1-\rho_2^2}.
\end{eqnarray}
In Eq.~(\ref{cond_y}),  $S(x,y,z;\bm{\theta})$ is unknown if the marginal distributions of variables $x$ and $z$ are undetermined. Due to that the variable $y$ obeys the Gaussian distribution, the highest probability of $y$ corresponds to $S(x,y,z;\bm{\theta})=0$.

\section{ The improved Amati correlation}\label{Sec_imp_Amati}

Now, we use the copula function introduced in the above section to investigate the correlation between $E_p$ and $E_{iso}$ of GRBs. After using  $x$, $y$ and $z$ to denote $\log\frac{E_p}{300\mathrm{keV}}$, $\log\frac{E_{iso}}{1\mathrm{erg}}$ and redshift $z$ of GRBs, respectively, the marginal distribution of $x$, $y$ and $z$ need to be given in order to determine $S(x,y,z;\bm{\theta})$. We  assume that
two Gaussian distributions for $x=\log\frac{E_p}{300\mathrm{keV}}$ and $y=\log \frac{E_{iso}}{1\mathrm{erg}}$ are
    \begin{eqnarray}\label{gauss_xy}
        f(x;\bar{a}_x,\sigma_x)= \frac{1}{\sqrt{2 \pi } \sigma_x} e^{-\frac{(\bar{a}_x-x)^2}{2 \sigma_x^2}}&,&~
        g(y;\bar{a}_y,\sigma_y)= \frac{1}{\sqrt{2 \pi } \sigma_y} e^{-\frac{(\bar{a}_y-y)^2}{2 \sigma_y^2}}.
    \end{eqnarray}
Here $\bar{a}$ and $\sigma$  represent the mean value and the standard deviation of the Gaussian distribution respectively.
And the CDFs of $x$, $y$, respectively,  have the forms
\begin{eqnarray}\label{CDF_xy} F(x;\bar{a}_x,\sigma_x)=\int_{-\infty}^{x}f(\tilde{x};\bar{a}_x,\sigma_x)d\tilde{x},~G(y;\bar{a}_y,\sigma_y)=\int_{-\infty}^{y}g(\tilde{y};\bar{a}_y,\sigma_y)d\tilde{y}.
\end{eqnarray}
We consider two different redshift distributions of GRB data:
	\begin{itemize}
		\item the first distribution is a special form with the PDF being $w(z)=z e^{-z}$~\citep{Wang2017}. Thus, the corresponding  CDF is:
		\begin{eqnarray}\label{power_z}
		 W(z)=1-e^{-z} (1+z).
		\end{eqnarray}
		\item the second CDF of GRB's redshift distribution is the empirical distribution~\citep{Dekking2005}:
		\begin{eqnarray}\label{empirical_z}
		W_N(z)= \left\{\begin{array}{ll}
		0 & \mbox{ if } z<z_1 \\
		i/n &\mbox{ if } z_i\le z< z_{i+1} \\
		1 &\mbox{ if } z_n\le z \ ,
		\end{array} \right.
		\end{eqnarray}
		where $\{z_1,z_2,\dots,z_n\}$ is an ordered list of redshifts of GRB data, which satisfies $z_1\le z_2\le\dots\le z_n$, and $n$ denotes the number of data.
	\end{itemize}

\begin{figure}
    \plotone{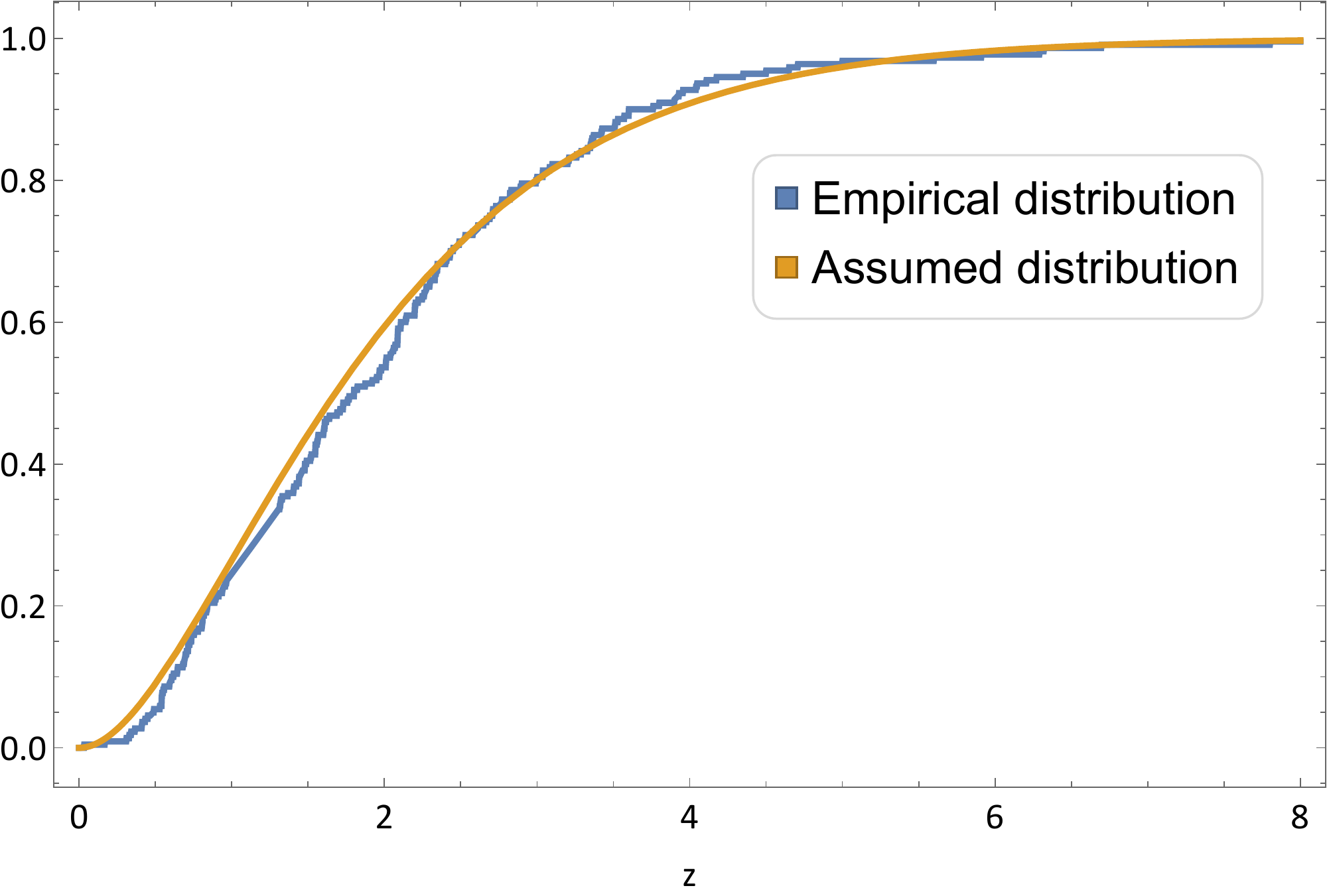}
    \caption{The CDFs of the empirical     and  assumed redshift distributions of GRB data. The maximum deviation between two CDFs is about 0.06 at $z\simeq2$, and is less than the critical value $D_{\alpha}\simeq0.09$, which means that the assumed distribution in our work  passes the K-S test.
        \label{Fig:KS_Test}}
\end{figure}

Since the first distribution of redshift $z$ of GRBs given in Eq.~(\ref{power_z}) is an assumed special form, we need to evaluate whether the redshift of observed GRB data obeys this form by using  the Kolmogorov-Smirnov test (K-S test). The K-S test bases on the distance measure $D$ which is defined to be
\begin{eqnarray}
D=\max |W_N(z)-W(z)|,
\end{eqnarray}
where $W_N(z)$ and $W(z)$ are the empirical  and   assumed CDFs of redshift distribution of GRBs respectively. Apparently, $D$ represents the maximum deviation between two CDFs.  In our K-S test, we set the significance level $\alpha $ to be $\alpha=0.05$, where $\alpha \equiv 1-P(\lambda_{\alpha})$ and $P(\lambda)$ is the
the Kolmogorov distribution
\begin{eqnarray}
P(\lambda)=\sum_{i=-\infty}^{+\infty}(-1)^{i}\exp\left(-2i^2\lambda^2\right)
\end{eqnarray}
obeyed by  the Kolmogorov stochasticity parameter $\lambda$. Here  $\lambda\equiv\sqrt{N}D$ and $N$  is the number of data.   Then,  a critical value $D_{\alpha}=\frac{\lambda_{\alpha}}{\sqrt{N}}\simeq0.09$ can be obtained, which means that if the data satisfies the assumed distribution, the probability that $D<D_{\alpha}$ is about $95\%$. Conversely, the probability that $D>D_{\alpha}$ is only $5\%$.
If $D>D_{\alpha}$, the assumed distribution will be rejected  since it is a rare event. Fig.~\ref{Fig:KS_Test} shows the CDFs of the empirical distribution $W_N(z)$ from the latest 220 GRB data points~\citep{Khadka2021} and the assumed distribution $W(z)$  used in our discussion. We find that $D\simeq0.06$ at $z\simeq 2$, which is less than $D_{\alpha}$.  Thus the assumed redshift distribution of GRB data  passes the K-S test and is acceptable.

Substituting the CDFs  given  in Eqs.~(\ref{CDF_xy},\ref{power_z},\ref{empirical_z}) 
 into Eqs.~(\ref{gauss_c}-\ref{cond_y}),  $S(x,y,z;\bm{\theta})$ can be derived. However, the concrete expression is very complicated, so we do not show it here.
From the equation of $S(x,y,z;\bm{\theta})=0$, we can obtain:
\begin{eqnarray}\label{3_Amati}
    y_{\mathrm{copula_1}}&=& a+b\, x+c\, \mathrm{erfc}^{-1}[2 W(z)],\nonumber\\
    y_{\mathrm{copula_2}}&=& a+b\, x+c\, \mathrm{erfc}^{-1}[2 W_N(z)].
\end{eqnarray}

Here, the subscripts $\mathrm{copula_1}$ and $\mathrm{copula_2}$ denote the correlation relations from the copula method with the assumed redshift distribution (Eq.~(\ref{power_z})) and the empirical redshift distribution (Eq. (\ref{empirical_z})), respectively,
 $\mathrm{erfc}$ is the complementary error function, and coefficients $a$, $b$ and $c$ are defined as
\begin{eqnarray}
    a&\equiv& \bar{a}_y-\frac{\left(\rho_2\rho_3-\rho_1\right)\bar{a}_x\sigma_y}{\left(\rho_2^2-1\right)\sigma_x},\nonumber\\
    b&\equiv& \frac{\left(\rho_2\rho_3-\rho_1\right)\sigma_y}{\left(\rho_2^2-1\right)\sigma_x},\nonumber\\
    c&\equiv& \frac{\sqrt{2}\sigma_y\left(\rho_3-\rho_1\rho_2\right)}{\rho_2^2-1}.
\end{eqnarray}
Eq.~(\ref{3_Amati}) are different from the standard  Amati correlation (Eq.~(\ref{2_Amati})) by a redshift-dependent term, and this redshift-dependent correction term is also different from that of the extended Amati correlation (Eq.~(\ref{exAmati})). We name these luminosity correlations from the Gaussian copula as   the improved  Amati correlations, and they are the main results of our paper.

\section{Hubble diagrams and GRB cosmology}\label{Sec_result}

 \subsection{The low-redshift calibration}
 To test the viability of the improved Amati correlations given in Eq.~(\ref{3_Amati}), we use two GRB data samples: one is the  latest GRB sample (A220)~\citep{Khadka2021}, which  contains 220 long GRBs  in the redshift range of $z\in[0.03,8.2]$, and the other is the higher-quality A118 data set~\citep{Khadka2020,Wang2016,Dirirsa2019} contained in A220 with redshift range of $z\in[0.34,8.2]$ since it has a tighter intrinsic scatter.
These GRBs are divided into the low-redshift part (79 and 20 GRBs at $z\in\left[0,1.4\right]$ in A220 and A118, respectively) and  the high-redshift one (141 and 98 GRBs at $z\in\left[1.4,8.2\right]$ in A220 and A118, respectively).
We will use the low-redshift GRB data of two data sets to determine the coefficients in $y_{\mathrm{copula}}$, and then extrapolate these results to the high redshift data to obtain their luminosity distances, which will then be used to constrain the cosmological model. Since the isotropic equivalent radiated energy $E_{iso}$ (see Eq.~(\ref{Eiso})) is dependent on the luminosity distance, a fiducial cosmological model needs to be chosen. Here, the spatially  flat $\Lambda$CDM with $\Omega_{\mathrm{m0}}=0.30$ and $H_0=70~ \mathrm{ km~s^{-1} Mpc^{-1}}$ is chosen as the fiducial model, where  $\Omega_{\mathrm{m0}}$ is the dimensionless present matter density parameter.  Then, the allowed regions of $a$, $b$, $c$ and $\sigma_{int}$ can be obtained by maximizing the D'Agostinis likelihood function
\begin{eqnarray}\label{Lc}
    \mathcal{L}(\sigma_{int},a,b,c)\propto\prod_{i=1}^{N} \frac{1}{\sqrt{\sigma_{int}^2+\sigma_{y,i}^2+b^2\sigma_{x,i}^2}}
    \times\exp\left[-\frac{[y_i-y_\mathrm{copula}(x_i,z_i; a, b, c)]^2}{2\left(\sigma_{int}^2+\sigma_{y,i}^2+b^2\sigma_{x,i}^2\right)}\right].
\end{eqnarray}
Here  $N=79$ or $20$, $\sigma_{int}$ is the intrinsic scatter  of GRBs, and $y_i$ denotes the observed $\log\frac{E_{iso}}{\mathrm{1erg}}$ of the low-redshift GRBs.   For a comparison,   the Amati and extended Amati correlations, which are denoted by  $y_{\mathrm{Amati}}$ and $y_\mathrm{exAmati}$ respectively, are also investigated.
\textbf{
}
In our analysis, the  {\it CosmoMC} code is used\footnote{The {\it CosmoMC} code is available at \href{https://cosmologist.info/cosmomc/}{https://cosmologist.info/cosmomc}.}.
The results are summarized  in Tab.~\ref{tab:param} and Tab.~\ref{tab:param_2}. 

From two tables,  one can see that in the case of the $\mathrm copula_1$ correlation relation the  value of intrinsic scatter $\sigma_{int}$ is always slightly smaller than the one from the Amati relation, which means that the quality of the correlation relation is improved slightly when  the $\mathrm copula_1$ correlation relation is used. 
A comparison of Tab.~\ref{tab:param} and Tab.~\ref{tab:param_2} reveals that the value of intrinsic scatter $\sigma_{int}$  from the A118 GRB data set is smaller than the one from the A220 GRB data set. Thus,  the 118 data set is a higher quality one compared to the A220 data set, which agrees with the results obtained in \citep{Khadka2021}. 
Furthermore, Tab.~\ref{tab:param} shows that the coefficient $c$ in $y_\mathrm{copula_1}$(or $y_\mathrm{copula_2}$) and coefficients $\alpha$ and $\beta$ in $y_\mathrm{exAmati}$  are at most $1\sigma$  away  from $0$, which  indicates that there is not strong support for nonzero values of these parameters from the A220 data set.  These results are consistent with what were obtained in \citep{Khadka2021} where the relation parameters $a$ and $b$ were found to be independent of redshift within the error bars. 
This character may also be found from  the Pearson linear correlation coefficients $\rho_1$, $\rho_2$ and $\rho_3$, which denote the degree of linear correlation between variables $x$, $y$ and $z$. We derive  $\rho_1=0.771$, $\rho_2=0.493$, $\rho_3=0.426$ from the 79 low-redshift GRBs in the A220 data set and $\rho_1=0.781$, $\rho_2=0.381$, $\rho_3=0.377$ from the 20 low-redshift GRBs in the A118 data set. Since $\rho_2$ and $\rho_3$ have   values about $0.4$ and they are smaller clearly than $\rho_1$, the linear correlations between $\log\frac{E_p}{300\mathrm{keV}}$ and $z$,  and $\log\frac{E_p}{300\mathrm{keV}}$ and $z$ are weak, and they are weaker than  the linear correlation between   $\log\frac{E_p}{300\mathrm{keV}}$ and $\log\frac{E_{iso}}{1\mathrm{erg}}$.

 To compare the correlation relations from copula function and the (extended) Amati relation, we compute the values of the Akaike information criterion (AIC)\citep{Akaike1974,Akaike1981} and the Bayesian information criterion (BIC)\citep{Schwarz1978}, which, respectively, are defined as:
\begin{eqnarray}\label{AIC&BIC}
\mathrm{AIC}&=&2p-2\ln(\mathcal{L}),\nonumber\\
\mathrm{BIC}&=&p\ln N-2\ln(\mathcal{L}),
\end{eqnarray}
where $\mathcal{L}$ is the likelihood function,  $p$ is the number of free parameters in a model and $N$ is the number of data.
Here we also compute the $\Delta $AIC($\Delta $BIC), which denotes the difference of AIC(BIC) with respect to the reference model (the standard Amati model here). $0<\Delta \mathrm{AIC}(\Delta \mathrm{BIC}) \le 2$ indicates difficulty in preferring a given model, $2<\Delta \mathrm{AIC}(\Delta \mathrm{BIC})\le 6$ means mild evidence against the given model, and $\Delta \mathrm{AIC}(\Delta \mathrm{BIC})>6$ suggests strong evidence against the model. The obtained values of $\Delta $AIC and $\Delta $BIC are summarized in Tabs.~\ref{tab:param} and \ref{tab:param_2}. We find that  except for the case of the extended Amati correlation   whose $\Delta\mathrm{AIC}$ given by the low-redshift GRBs in the A118 data set is larger than $2$, the  AIC  cannot determine the model preferred by the data, while the  BIC indicates that the standard Amati correlation remains to be favored mildly since it has the least model parameters.

\begin{deluxetable}{cccccccccccc}
    \tablenum{1}
    \tablecaption{\label{tab:param}}
    \tablewidth{0pt}
    \tablehead{
        & \multicolumn{2}{c}{Amati} & & \multicolumn{2}{c}{extended Amati} & & \multicolumn{2}{c}{$\mathrm{copula_1}$} & & \multicolumn{2}{c}{$\mathrm{copula_2}$}\\
        \cline{1-3} \cline{5-6} \cline{8-9} \cline{11-12}
        \colhead{} &\colhead{\it Best-fit($\sigma$)} &\colhead{\it 0.68 CL} &\colhead{} &\colhead{\it Best-fit($\sigma$)} &\colhead{\it 0.68 CL} &\colhead{} &\colhead{\it Best-fit($\sigma$)} &\colhead{\it 0.68 CL} &\colhead{} &\colhead{\it Best-fit($\sigma$)} &\colhead{\it 0.68 CL}
    }
    \startdata
    $\sigma_{int}$ & $0.512(0.045)$ & ${}^{+0.054}_{-0.034}$ & & $0.503(0.046)$ & ${}^{+0.063}_{-0.025}$ & & $0.510(0.045)$ & ${}^{+0.054}_{-0.033}$ & & $0.509(0.045)$ & ${}^{+0.060}_{-0.029}$ \\
    $a$ & $52.710(0.061)$ & ${}^{+0.060}_{-0.061}$ & & $52.587(0.333)$ & ${}^{+0.324}_{-0.334}$ & & $52.847(0.144)$ & ${}^{+0.145}_{-0.137}$ & & $52.812(0.149)$ & ${}^{+0.141}_{-0.156}$ \\
    $b$ & $1.290(0.126)$ & ${}^{+0.126}_{-0.123}$ & & $1.521(0.367)$ & ${}^{+0.422}_{-0.301}$ & & $1.209(0.150)$ & ${}^{+0.145}_{-0.148}$ & & $1.231(0.145)$ & ${}^{+0.156}_{-0.128}$ \\
    $c$ & $-$ & ${-}^{}_{}$ & & $-$ & ${-}^{}_{}$ & & $-0.217(0.207)$ & ${}^{+0.208}_{-0.198}$ & & $-0.142(0.192)$ & ${}^{+0.200}_{-0.180}$ \\
    $\alpha$ & $-$ & ${-}^{}_{}$ & & $0.319(0.729)$ & ${}^{+0.734}_{-0.702}$ & & $-$ & ${-}^{}_{}$ & & $-$ & ${-}^{}_{}$ \\
    $\beta$ & $-$ & ${-}^{}_{}$ & & $-0.680(0.779)$ & ${}^{+0.624}_{-0.915}$ & & $-$ & ${-}^{}_{}$ & & $-$ & ${-}^{}_{}$ \\
    \hline
    $-2\ln\mathcal{L}$ & \multicolumn{2}{c}{121.514} & & \multicolumn{2}{c}{119.492} & & \multicolumn{2}{c}{120.281} & & \multicolumn{2}{c}{121.012}\\
   $\Delta$AIC & \multicolumn{2}{c}{-} & & \multicolumn{2}{c}{1.978} & & \multicolumn{2}{c}{0.767} & & \multicolumn{2}{c}{1.498}\\
   $\Delta$BIC & \multicolumn{2}{c}{-} & & \multicolumn{2}{c}{6.717} & & \multicolumn{2}{c}{3.136} & & \multicolumn{2}{c}{3.867}\\
    \enddata
    \tablecomments{The best-fitted values, standard deviations, and the 68\% confidence level (CL) of coefficients of $y_\mathrm{Amati}$, $y_\mathrm{exAmati}$, $y_\mathrm{copula_1}$ and $y_\mathrm{copula_2}$  from the 79 low-redshift ($z<1.4$) long GRBs in A220 data set. Here $\Delta \mathrm{AIC}(\Delta \mathrm{BIC})$ denotes the difference of AIC(BIC) with standard Amati model.
    }
\end{deluxetable}

\begin{deluxetable}{cccccccccccc}
	\tablenum{2}
	\tablecaption{\label{tab:param_2}}
	\tablewidth{0pt}
	\tablehead{
		& \multicolumn{2}{c}{Amati} & & \multicolumn{2}{c}{extended Amati} & & \multicolumn{2}{c}{$\mathrm{copula_1}$} & & \multicolumn{2}{c}{$\mathrm{copula_2}$}\\
		\cline{1-3} \cline{5-6} \cline{8-9} \cline{11-12}
		\colhead{} &\colhead{\it Best-fit($\sigma$)} &\colhead{\it 0.68 CL} &\colhead{} &\colhead{\it Best-fit($\sigma$)} &\colhead{\it 0.68 CL} &\colhead{} &\colhead{\it Best-fit($\sigma$)} &\colhead{\it 0.68 CL} &\colhead{} &\colhead{\it Best-fit($\sigma$)} &\colhead{\it 0.68 CL}
	}
	\startdata
	$\sigma_{int}$ & $0.429(0.089)$ & ${}^{+0.117}_{-0.044}$ & & $0.438(0.099)$ & ${}^{+0.135}_{-0.041}$ & & $0.418(0.098)$ & ${}^{+0.149}_{-0.025}$ & & $0.430(0.097)$ & ${}^{+0.137}_{-0.035}$ \\
	$a$ & $52.860(0.111)$ & ${}^{+0.105}_{-0.110}$ & & $52.730(0.609)$ & ${}^{+0.534}_{-0.6496}$ & & $52.9586(0.282)$ & ${}^{+0.281}_{-0.257}$ & & $52.935(0.282)$ & ${}^{+0.270}_{-0.266}$ \\
	$b$ & $0.990(0.205)$ & ${}^{+0.210}_{-0.187}$ & & $1.525(1.158)$ & ${}^{+1.230}_{-1.168}$ & & $0.967(0.231)$ & ${}^{+0.213}_{-0.229}$ & & $0.976(0.235)$ & ${}^{+0.218}_{-0.230}$ \\
	$c$ & $-$ & ${-}^{}_{}$ & & $-$ & ${-}^{}_{}$ & & $-0.170(0.425)$ & ${}^{+0.394}_{-0.423}$ & & $-0.103(0.368)$ & ${}^{+0.339}_{-0.360}$ \\
	$\alpha$ & $-$ & ${-}^{}_{}$ & & $0.329(1.327)$ & ${}^{+1.405}_{-1.170}$ & & $-$ & ${-}^{}_{}$ & & $-$ & ${-}^{}_{}$ \\
	$\beta$ & $-$ & ${-}^{}_{}$ & & $-1.136(2.270)$ & ${}^{+2.362}_{-2.476}$ & & $-$ & ${-}^{}_{}$ & & $-$ & ${-}^{}_{}$ \\
	\hline
	$-2\ln\mathcal{L}$ & \multicolumn{2}{c}{23.374} & & \multicolumn{2}{c}{22.938} & & \multicolumn{2}{c}{23.119} & & \multicolumn{2}{c}{23.240}\\
	$\Delta$AIC & \multicolumn{2}{c}{-} & & \multicolumn{2}{c}{3.564} & & \multicolumn{2}{c}{1.745} & & \multicolumn{2}{c}{1.866}\\
	$\Delta$BIC & \multicolumn{2}{c}{-} & & \multicolumn{2}{c}{5.555} & & \multicolumn{2}{c}{2.741} & & \multicolumn{2}{c}{2.862}\\
	\enddata
	\tablecomments{The best-fitted values, standard deviations, and the 68\% confidence level (CL) of coefficients of $y_\mathrm{Amati}$, $y_\mathrm{exAmati}$, $y_\mathrm{copula_1}$ and $y_\mathrm{copula_2}$  from the 20 low-redshift ($z<1.4$) long GRBs in A118 data set.
	}
\end{deluxetable}

\begin{figure}
 \includegraphics[width=0.5\textwidth]{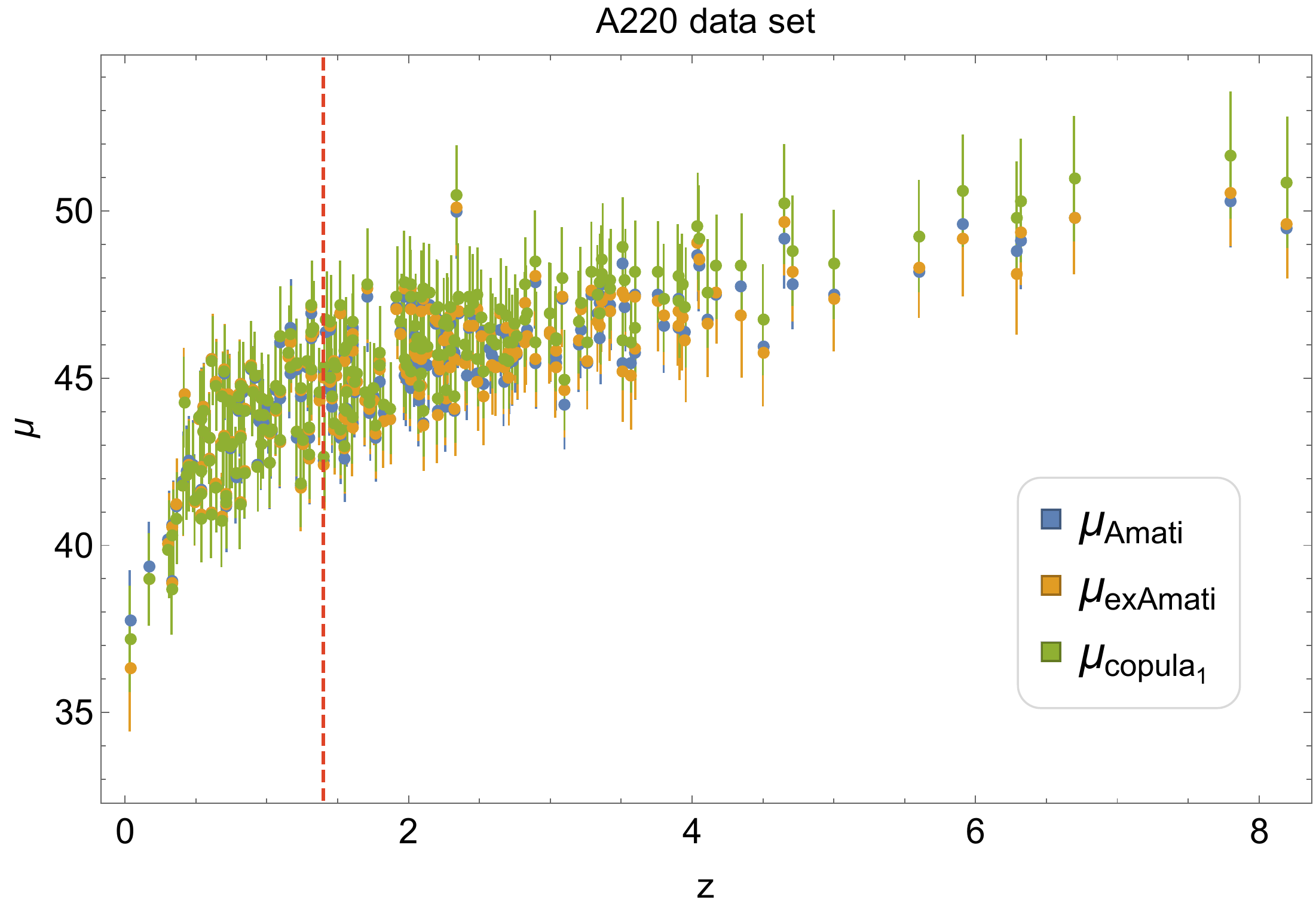}
    \includegraphics[width=0.5\textwidth]{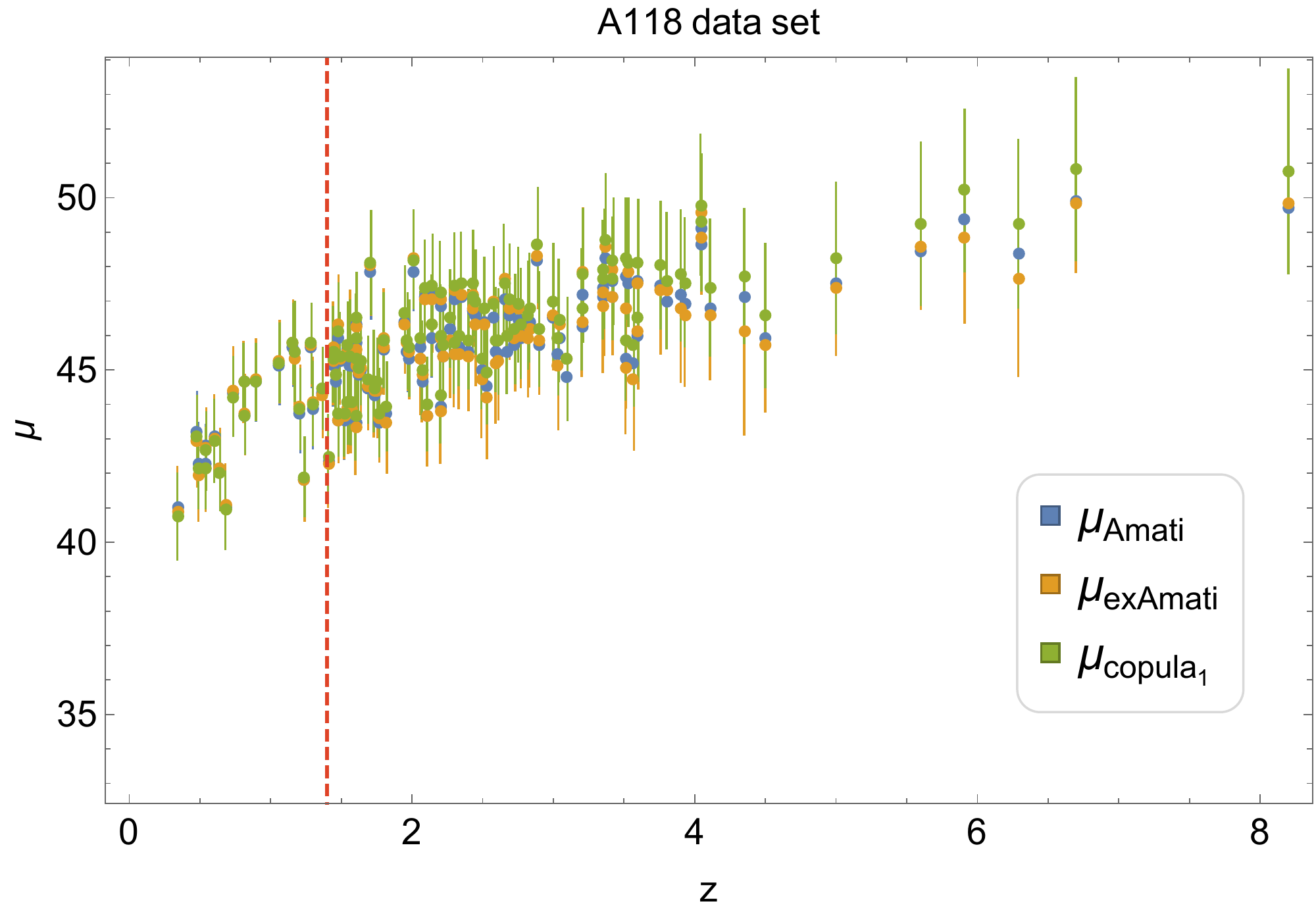}
    \caption{The Hubble diagrams of 220 and 118 long GRBs calibrated from three different correlations ($y_\mathrm{Amati}$, $y_\mathrm{exAmati}$ and $y_\mathrm{copula_1}$).   The red dashed line denotes $z=1.4$.
    \label{Fig:Hubble_dia}}
\end{figure}

Extrapolating directly the values of the coefficients in Tab.~\ref{tab:param} and Tab.~\ref{tab:param_2} from the low-redshift GRB data to the high-redshift samples we can construct the Hubble diagram of GRBs.
The distance modulus of GRBs and their errors  are obtained from Eqs.~(\ref{mu}) and (\ref{mu_err}). As an example, in Fig.~\ref{Fig:Hubble_dia}, we show the Hubble diagrams of 220 and 118 long GRBs obtained from three different correlations ($y_\mathrm{Amati}$, $y_\mathrm{exAmati}$ and $y_\mathrm{copula_1}$), respectively. 
Here we do not plot the Hubble diagram based on the $y_\mathrm{copula_2}$ correlation since it is very similar to the one of $y_\mathrm{copula_1}$.
The values of distance modulus obtained from the copula method are less at low-redshift ($z\lesssim1$) regions and larger at high-redshift ($z\gtrsim1$) regions than the ones from the standard Amati correlation because a correction exists in the right hand side of Eq.~(\ref{3_Amati}).

To test whether the GRB can be regarded as the viable cosmological indicator, we can constrain the  $\Lambda$CDM model from the distance modulus of GRBs obtained in the above subsection, and check whether $\Omega_\mathrm{m0}=0.3$ is allowed after setting $H_0=70~\mathrm{km~s^{-1}Mpc^{-1}}$, which is used in calibrating the improved Amati correlations from the low redshift GRBs.  The  method  of minimizing $\chi^2$ is used to constrain  $\Omega_{\mathrm {m0}}$.
We  consider two different samples for each data set: the 141 high-redshift GRBs  and the total 220 GRBs for A220, and the 98 high-redshift GRBs  and the total 118 GRBs for A118. 
We must emphasize here that since  $W_N(8.2)=1$ when the maximum redshift GRB data ($z=8.2$) is considered,   $\mathrm{erfc}^{-1}[2W_N(8.2)]=-\infty$ in $y_{copula_2}$  and thus  the data point with $z=8.2$ will be ignored when the $y_{copula_2}$ is used to constrain   $\Omega_{\mathrm {m0}}$.

\begin{figure}
    \includegraphics[width=0.5\textwidth]{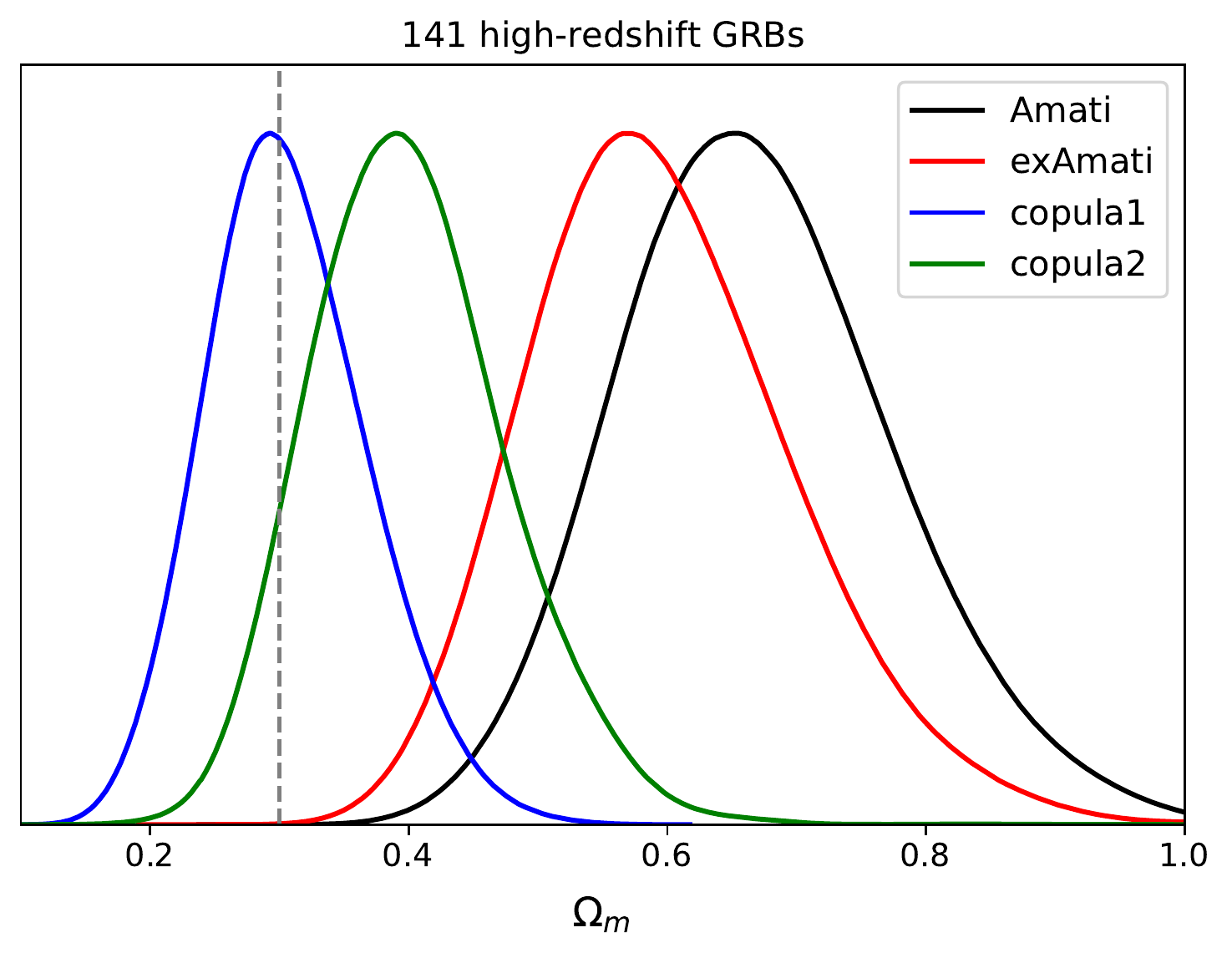}
    \includegraphics[width=0.5\textwidth]{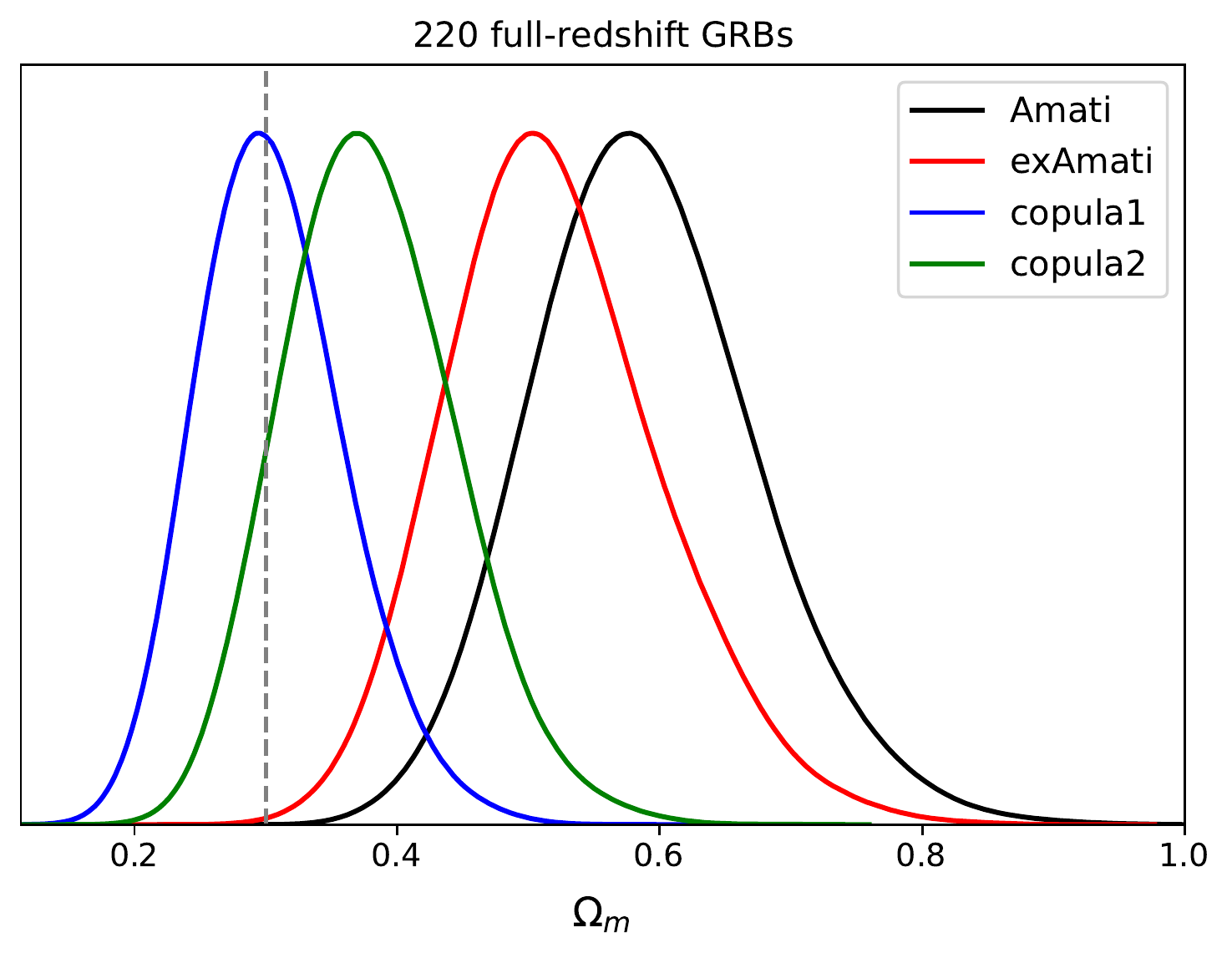}
    \caption{The probability density plots of  $\Omega_\mathrm{m0}$ in the $\Lambda$CDM model with $H_0=70$ $\mathrm{km}~\mathrm{s^{-1} Mpc^{-1}}$.
        The left and right panels show the results from the 141 high-redshift GRBs and full-redshift 220 GRBs, respectively.
        The gray dashed line denotes the $\Omega_{\mathrm{m0}}=0.3$ which is the value of fiducial model.
        \label{Fig:omegaplot}}
\end{figure}

\begin{figure}
	\includegraphics[width=0.5\textwidth]{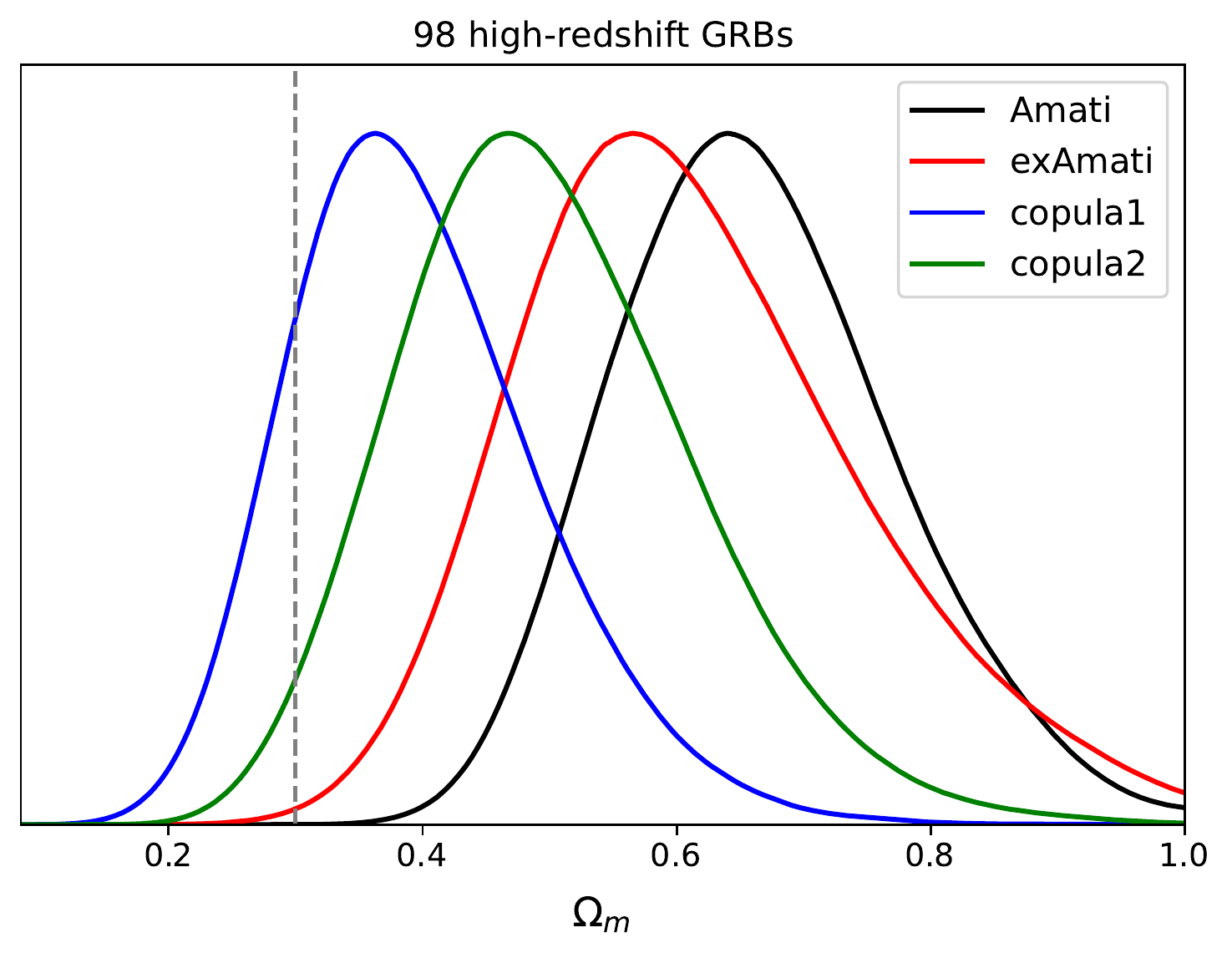}
	\includegraphics[width=0.5\textwidth]{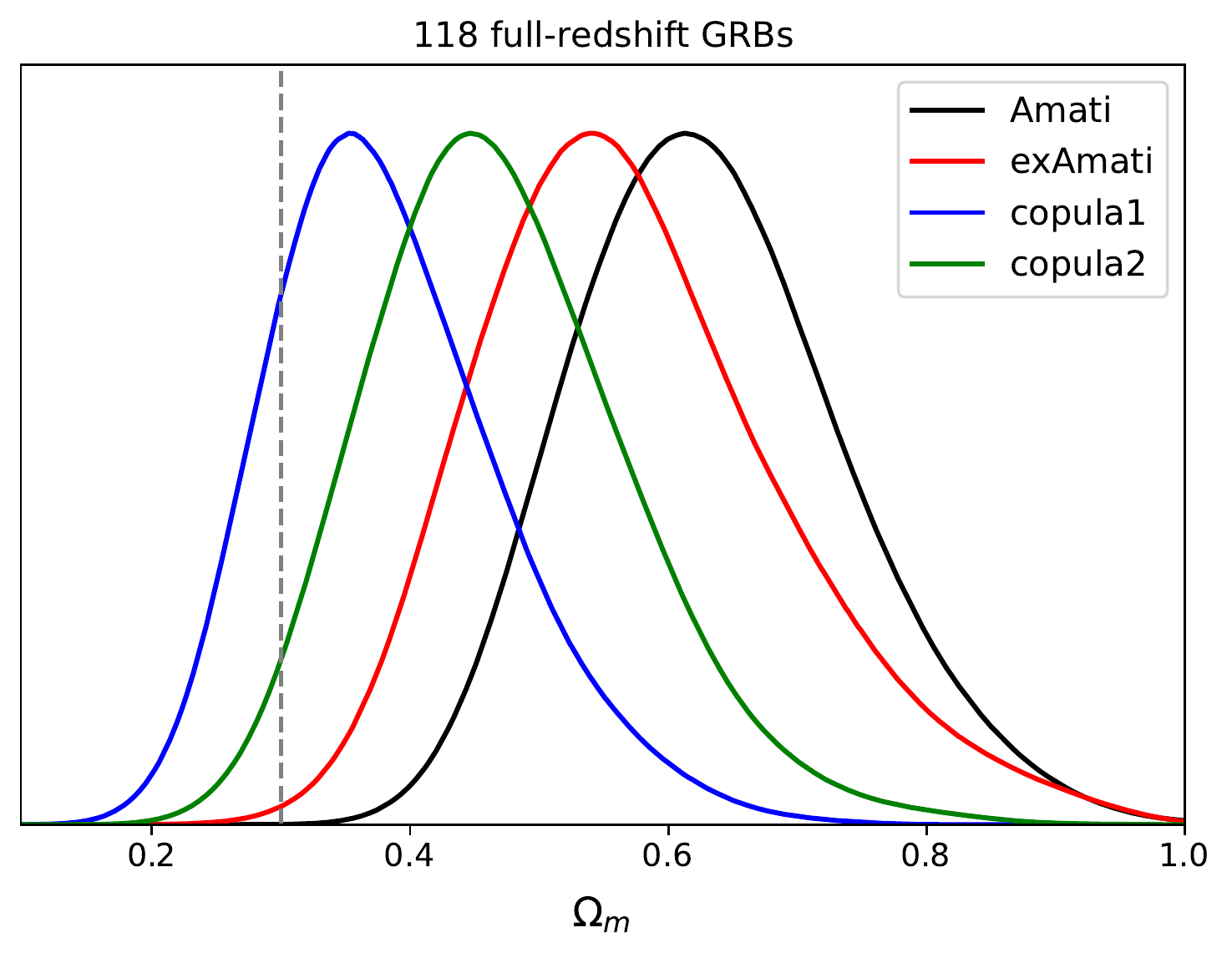}
	\caption{The probability density plots of  $\Omega_\mathrm{m0}$ in the $\Lambda$CDM model with $H_0=70$ $\mathrm{km}~\mathrm{s^{-1} Mpc^{-1}}$.
		The left and right panels show the results from the 98 high-redshift GRBs and full-redshift 118 GRBs, respectively.
		The gray dashed line denotes the $\Omega_{\mathrm{m0}}=0.3$ which is the value of fiducial model.
		\label{Fig:omegaplot_118}}
\end{figure}

The probability density plots of $\Omega_{\mathrm {m0}}$ for two data sets are shown in Fig. \ref{Fig:omegaplot} and Fig. \ref{Fig:omegaplot_118}, and the best-fitted values with the standard deviation and  $68\%$ confidence level (CL) are summarized in Tab.~\ref{tab:LCDM} and Tab.~\ref{tab:LCDM_118}, respectively. 
It is easy to see that for all data sets  the results from the standard Amati correlation and the extended Amati correlation deviate apparently from $\Omega_{\mathrm {m0}}=0.3$.  The results from improved Amati correlation based on  an empirical distribution of redshift ($y_\mathrm{copula_2}$) are better than the ones from the (extended) Amati correlation although they are $1\sigma$ away from $\Omega_{\mathrm {m0}}=0.3$, while those from $y_\mathrm{copula_1}$ are always consistent with $\Omega_{\mathrm {m0}}=0.3$ at the $1\sigma$ confidence level. Apparently, the  results from  $y_\mathrm{copula_2}$ are  not as good as those from  $y_\mathrm{copula_1}$. This is attributed to that the number of GRBs is still inadequate to construct the empirical distribution precisely. 
Comparing the constraint on $\Omega_{\mathrm{m0}}$ from the high-redshift and full-redshift GRB data shown in  Tabs.~\ref{tab:LCDM} and \ref{tab:LCDM_118}, we find that the values of $\Omega_{\mathrm{m0}}$ from the full-redshift data are closer to $0.3$ than those from the high-redshift data. This is because the low-redshift data are calibrated with $\Omega_{\mathrm{m0}}=0.3$. When changing the GRB data from the high-redshift region to the full-redshift one,  the A220 data give  the maximum variation of $\Omega_{\mathrm{m0}}$ for the case of the extended Amati correlation, which is about $12\%$. However, this variation is very small at  $y_\mathrm{copula_1}$ and  $y_\mathrm{copula_2}$ cases. From Tabs.~\ref{tab:LCDM} and \ref{tab:LCDM_118} and Figs.~\ref{Fig:omegaplot} and \ref{Fig:omegaplot_118}, one can also see that  the values of $\Omega_{\mathrm {m0}}$ from the total A220 data set are much closer to $0.3$ than  those  from  the full-redshift  A118 data set. This  is  attributed to that the ratio of the number of the calibrated to the uncalibrated GRBs  in the A220 data set, which is  about 0.56 (79/141),   is apparently larger than that in the A118 data set, which is only 0.20 (20/98).  In addition, It is easy to find  that our constraints on $\Omega_{\mathrm{m0}}$ from  the A118 and A220 data samples differ very significantly from what were obtained in \citep{Khadka2021} where the A118 and A220 data limit  $\Omega_{\mathrm{m0}}>0.230$ and $>0.455$, respectively,  at the $95\%$ CL in the $\Lambda$CDM model. This difference originates from that  we utilize the low-redshift calibration method while the  simultaneous fitting was  used  in  \citep{Khadka2021}.

\begin{deluxetable}{cccccccc}
    \tablenum{3}
    \tablecaption{\label{tab:LCDM}}
    \tablewidth{0pt}
    \tablehead{
         & \multicolumn{3}{c}{high-redshift} & & \multicolumn{3}{c}{full-redshift} \\
        \cline{2-4} \cline{6-8}
         &\colhead{$\Omega_{\mathrm{m0}}(\sigma)$} & \colhead{68\%CL} &\colhead{$\chi^2$}& &\colhead{$\Omega_{\mathrm{m0}}(\sigma)$} & \colhead{68\%CL} &\colhead{$\chi^2$}
    }
    \startdata
    Amati &$0.649(0.106)$ & ${}^{+0.114}_{-0.096}$ & 90.535 & & $0.589(0.086)$ & ${}^{+0.091}_{-0.077}$ & 168.851 \\
    \hline
    extend Amati &$0.574(0.103)$ & ${}^{+0.107}_{-0.093}$ & 86.194 & & $0.507(0.083)$ & ${}^{+0.083}_{-0.078}$ & 164.098\\
    \hline
    $\mathrm{copula_1}$ &$0.295(0.063)$ & ${}^{+0.067}_{-0.057}$ & 86.050 & & $0.296(0.057)$ & ${}^{+0.058}_{-0.053}$ & 160.779\\
    \hline
    $\mathrm{copula_2}$ &$0.385(0.077)$ & ${}^{+0.080}_{-0.068}$ & 84.264 & & $0.368(0.067)$ & ${}^{+0.071}_{-0.060}$ & 159.941
    \enddata
    \tablecomments{The best-fitted value of   $\Omega_{\mathrm {m0}}$ with the standard deviation $\sigma$ and the $68\%$ CL. The results are obtained from A220 data set.}
\end{deluxetable}

\begin{deluxetable}{cccccccc}
	\tablenum{4}
	\tablecaption{\label{tab:LCDM_118}}
	\tablewidth{0pt}
	\tablehead{
		& \multicolumn{3}{c}{high-redshift} & & \multicolumn{3}{c}{full-redshift} \\
		\cline{2-4} \cline{6-8}
		&\colhead{$\Omega_{\mathrm{m0}}(\sigma)$} & \colhead{68\%CL} &\colhead{$\chi^2$}& &\colhead{$\Omega_{\mathrm{m0}}(\sigma)$} & \colhead{68\%CL} &\colhead{$\chi^2$}
	}
	\startdata
	Amati &$0.638(0.109)$ & ${}^{+0.119}_{-0.100}$ & 69.710 & & $0.607(0.105)$ & ${}^{+0.116}_{-0.094}$ & 88.551 \\
	\hline
	extend Amati &$0.579(0.133)$ & ${}^{+0.138}_{-0.123}$ & 50.431 & & $0.540(0.120)$ & ${}^{+0.123}_{-0.108}$ & 66.905\\
	\hline
	$\mathrm{copula_1}$ &$0.390(0.101)$ & ${}^{+0.105}_{-0.086}$ & 46.979 & & $0.381(0.094)$ & ${}^{+0.099}_{-0.081}$ & 64.609\\
	\hline
	$\mathrm{copula_2}$ &$0.466(0.117)$ & ${}^{+0.130}_{-0.097}$ & 46.563 & & $0.445(0.103)$ & ${}^{+0.110}_{-0.090}$ & 63.794
	\enddata
	\tablecomments{The best-fitted value of   $\Omega_{\mathrm {m0}}$ with the standard deviation $\sigma$ and the $68\%$ CL. The results are obtained from A118 data set.}
\end{deluxetable}

\subsection{The simultaneous fitting}
To show clearly the difference of the results from the GRB low-redshift calibration  and the simultaneous fitting, now we follow the steps as in \citep{Khadka2021,Cao2022} to use the simultaneous fitting method to constrain the cosmological parameter  and the coefficients of four correlation relations. After setting $H_0=70~\mathrm{km~s^{-1}Mpc^{-1}}$, the constraints on $\Omega_{\mathrm{m0}}$ and the  coefficients  of four correlation relations can be obtained from the A118 and A220 data samples by   using the D'Agostinis likelihood function (Eq.~\ref{Lc}).

\begin{figure}
	\includegraphics[width=0.52\textwidth]{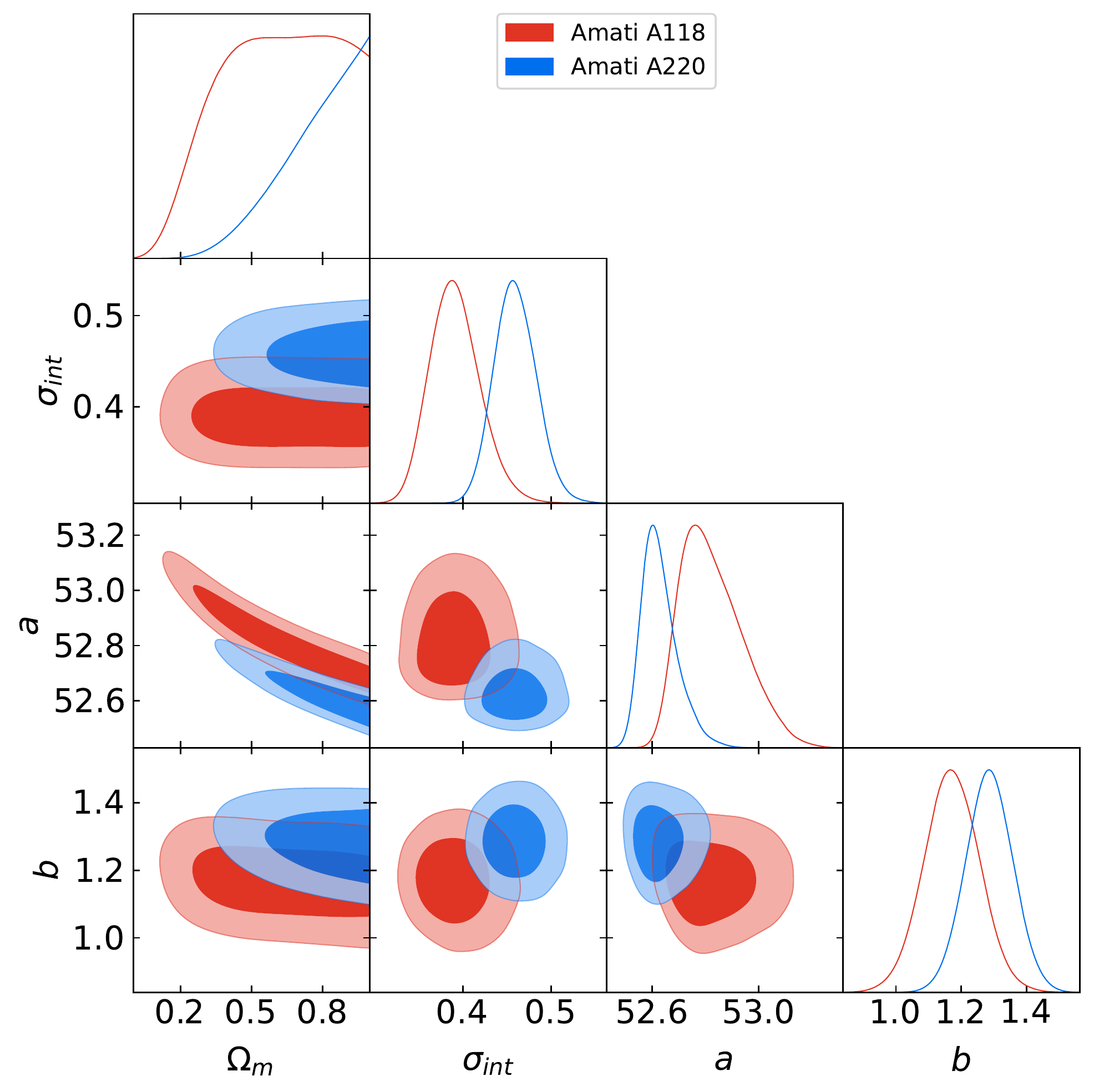}
	\includegraphics[width=0.52\textwidth]{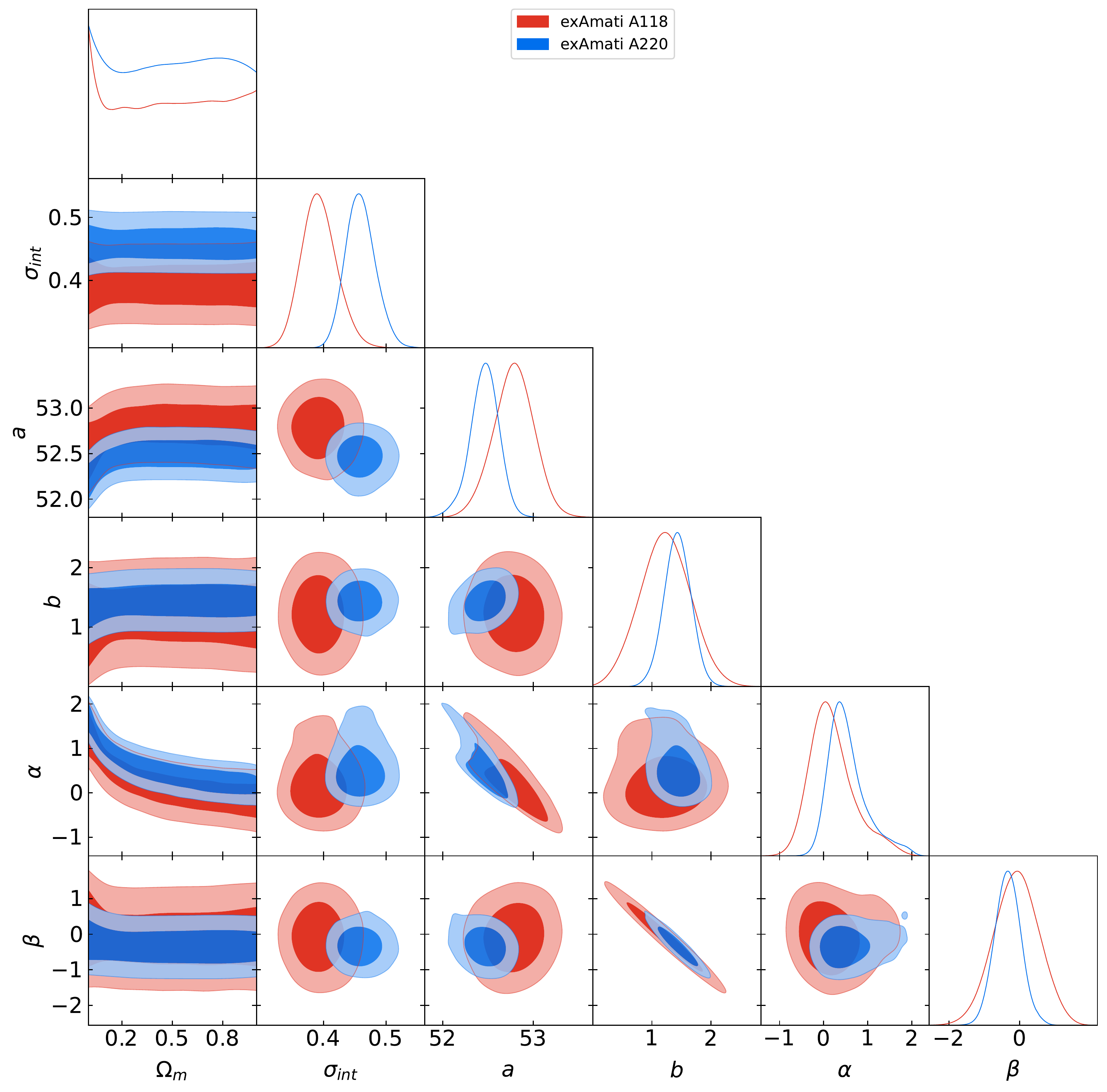}
	\includegraphics[width=0.52\textwidth]{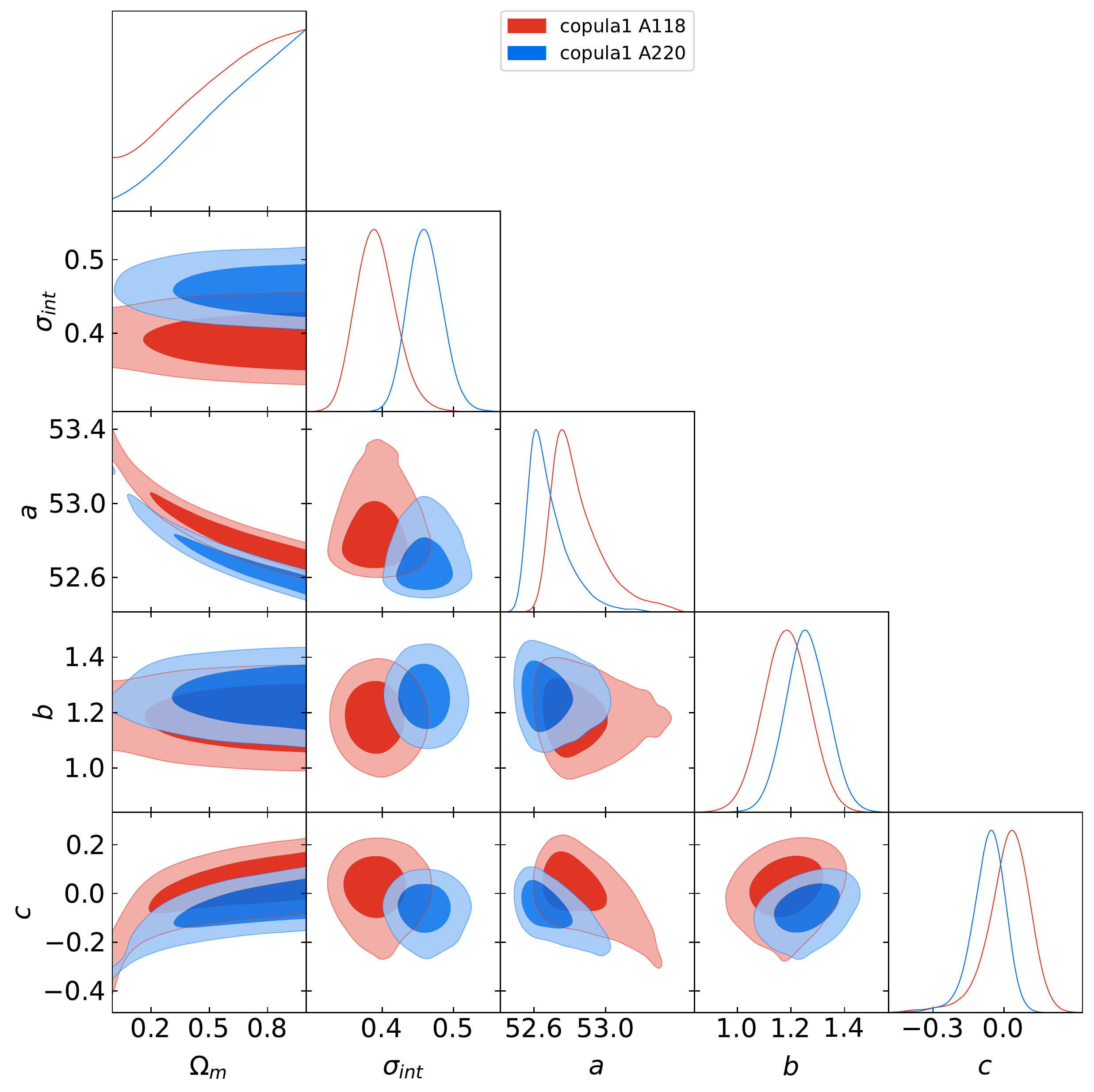}
	\includegraphics[width=0.52\textwidth]{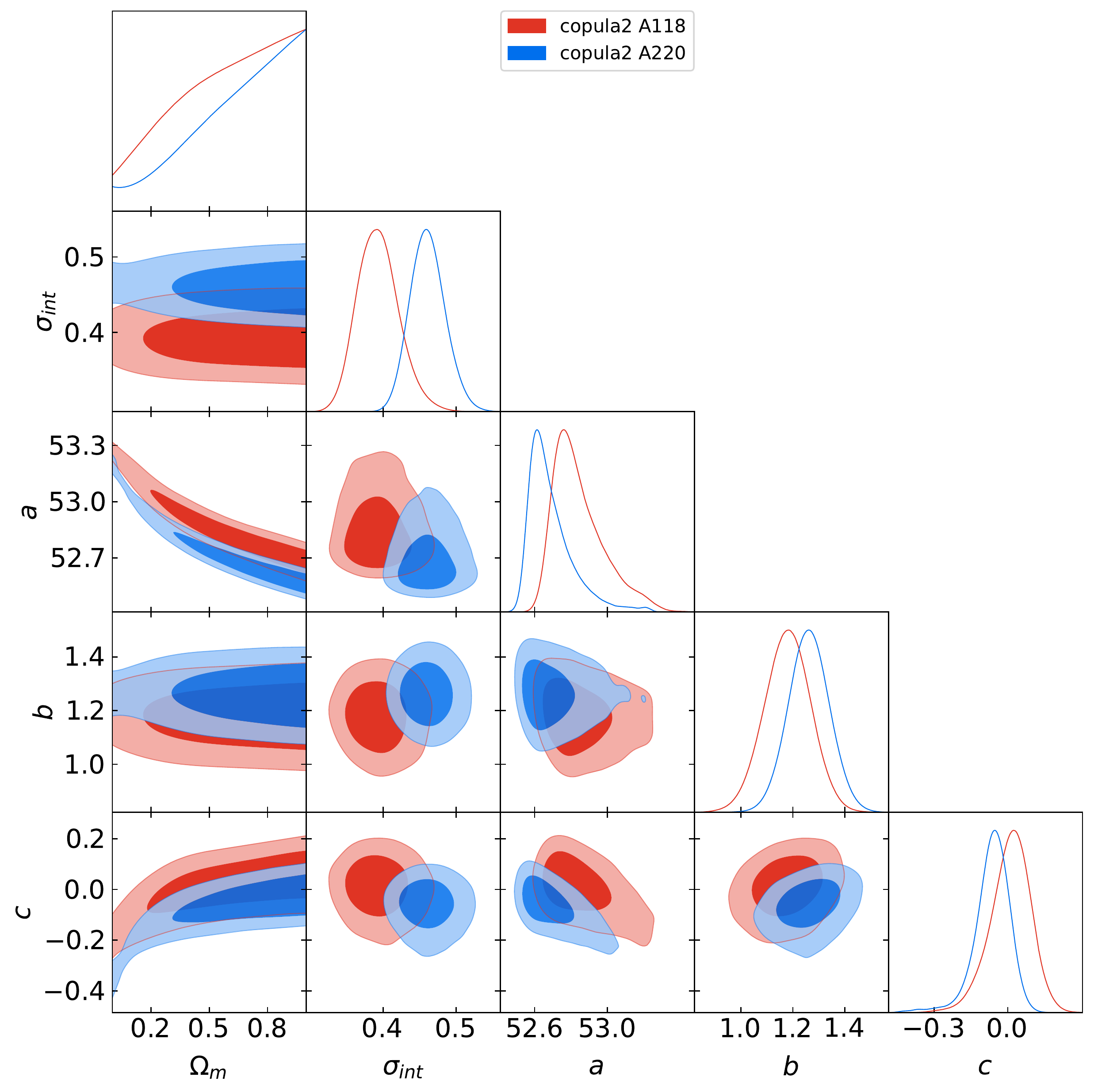}
	\caption{
		The results of simultaneously fitting the flat $\Lambda$CDM model and four correlations via the A118 (red) and A220 (blue) data sets. The upper left, upper right, lower left and lower right panels denotes the Amati, extended Amati, $\mathrm{copula_1}$ and $\mathrm{copula_2}$ correlations, respectively.
		\label{Fig:Simultaneous_Omega}}
\end{figure}

Fig. \ref{Fig:Simultaneous_Omega} shows  one dimensional probability density plots and contour plots of $\Omega_{\mathrm{m0}}$ and the coefficients of four correlation relations, and the marginalized mean values with the standard deviation and the $68\%$ CL are summarized in Tab. \ref{tab:simultaneous_A220}.
We find that the results from  the standard Amati correlation are  slightly different from what were obtained in \citep{Cao2022}.  This is because the peak energy of the GRB 081121 data point, which is released in \citep{Dirirsa2019} and used in  \citep{Cao2022},  is different  from the one given in Tab. 1 of \citep{Amati2009}, and it actually corresponds to the distance modulus rather than the peak energy in Tab. 4 of \cite{Wang2016}). Thus, there is an error in the peak energy of the GRB 081121 used in  \citep{Cao2022}. If this error were not corrected, we would obtain the same result as \cite{Cao2022}.  In our analysis we have corrected this error. Comparing Tab. \ref{tab:simultaneous_A220} and Tabs. \ref{tab:param} and \ref{tab:param_2}, we find that the values of $c$ from the simultaneous fitting are smaller and closer to zero than the ones  from the low-redshift calibration. This seems to indicate that the evolutionary character with redshift of Amati correlation becomes weaker when  more high redshift data are used.  Fig. \ref{Fig:Simultaneous_Omega}  and Tab. \ref{tab:simultaneous_A220} show that when the simultaneous fitting method is used,  the results from the improved Amati correlations are similar to those from the standard Amati correlation since the GRB data  favor a large value of $\Omega_{\mathrm{m0}}$ and can only give a low bound limit on $\Omega_{\mathrm{m0}}$ although the low bound limits from the improved Amati correlations are smaller clearly than the one from the standard Amati correlation.  If the extended Amati correlation is considered, there is almost no constraint on $\Omega_{\mathrm{m0}}$  from GRB, which should be attributed to that this relation has  more coefficients. Thus, once the simultaneous fitting method is used, we can not find that the improve Amati correlations obtained in this paper are apparently better  than the standard and extended Amati correlations.  These results are different significantly from those obtained in the  subsection above with the  low-redshift calibration method.

 \begin{deluxetable}{ccccccccccccc}
	\tablenum{5}
	\tablecaption{\label{tab:simultaneous_A220}}
	\tablewidth{0pt}
	\tablehead{
		Data set& & \multicolumn{2}{c}{Amati} & & \multicolumn{2}{c}{extended Amati} & & \multicolumn{2}{c}{$\mathrm{copula_1}$} & & \multicolumn{2}{c}{$\mathrm{copula_2}$}\\
		\cline{3-4} \cline{6-7} \cline{9-10} \cline{12-13}
		&\colhead{} &\colhead{\it Mean($\sigma$)} &\colhead{\it 0.68 CL} &\colhead{} &\colhead{\it Mean($\sigma$)} &\colhead{\it 0.68 CL} &\colhead{} &\colhead{\it Mean($\sigma$)} &\colhead{\it 0.68 CL} &\colhead{} &\colhead{\it Mean($\sigma$)} &\colhead{\it 0.68 CL}
	}
	\startdata
	&$\Omega_{\mathrm{m0}}$ & $>0.713$ & ${-}^{}_{}$ & & $-$ & ${-}^{}_{}$ & & $>0.554$ & ${-}^{}_{}$ & & $>0.550$ & ${-}^{}_{}$ \\
	&$\sigma_{int}$ & $0.459(0.024)$ & ${}^{+0.022}_{-0.025}$ & & $0.459(0.0.024)$ & ${}^{+0.021}_{-0.026}$ & & $0.460(0.024)$ & ${}^{+0.022}_{-0.026}$ & & $0.461(0.024)$ & ${}^{+0.022}_{-0.026}$ \\
	&$a$ & $52.630(0.068)$ & ${}^{+0.045}_{-0.079}$ & & $52.462(0.158)$ & ${}^{+0.160}_{-0.140}$ & & $52.685(0.118)$ & ${}^{+0.052}_{-0.130}$ & & $52.692(0.127)$ & ${}^{+0.048}_{-0.140}$ \\
	A220&$b$ & $1.286(0.072)$ & ${}^{+0.072}_{-0.072}$ & & $1.434(0.228)$ & ${}^{+0.230}_{-0.230}$ & & $1.258(0.078)$ & ${}^{+0.078}_{-0.078}$ & & $1.262(0.079)$ & ${}^{+0.078}_{-0.078}$ \\
	&$c$ & $-$ & ${-}^{}_{}$ & & $-$ & ${-}^{}_{}$ & & $-0.066(0.074)$ & ${}^{+0.077}_{-0.057}$ & & $-0.064(0.075)$ & ${}^{+0.076}_{-0.052}$ \\
	&$\alpha$ & $-$ & ${-}^{}_{}$ & & $0.544(0.447)$ & ${}^{+0.240}_{-0.510}$ & & $-$ & ${-}^{}_{}$ & & $-$ & ${-}^{}_{}$ \\
	&$\beta$ & $-$ & ${-}^{}_{}$ & & $-0.337(0.370)$ & ${}^{+0.370}_{-0.370}$ & & $-$ & ${-}^{}_{}$ & & $-$ & ${-}^{}_{}$ \\
	\hline
	&$\Omega_{\mathrm{m0}}$ & $0.610(0.230)$ & ${}^{+0.340}_{-0.170}$ & & $<0.680$ & ${-}^{}_{}$ & & $>0.470$ & ${-}^{}_{}$ & & $>0.459$ & ${-}^{}_{}$ \\
	&$\sigma_{int}$ & $0.391(0.028)$ & ${}^{+0.024}_{-0.031}$ & & $0.394(0.028)$ & ${}^{+0.024}_{-0.031}$ & & $0.392(0.028)$ & ${}^{+0.025}_{-0.031}$ & & $0.394(0.029)$ & ${}^{+0.025}_{-0.031}$ \\
	&$a$ & $52.826(0.115)$ & ${}^{+0.082}_{-0.140}$ & & $52.786(0.228)$ & ${}^{+0.240}_{-0.210}$ & & $52.846(0.147)$ & ${}^{+0.067}_{-0.170}$ & & $52.845(0.142)$ & ${}^{+0.075}_{-0.170}$ \\
	A118&$b$ & $1.171(0.085)$ & ${}^{+0.085}_{-0.085}$ & & $1.234(0.422)$ & ${}^{+0.418}_{-0.420}$ & & $1.183(0.086)$ & ${}^{+0.086}_{-0.086}$ & & $1.179(0.089)$ & ${}^{+0.089}_{-0.089}$ \\
	&$c$ & $-$ & ${-}^{}_{}$ & & $-$ & ${-}^{}_{}$ & & $0.017(0.097)$ & ${}^{+0.100}_{-0.068}$ & & $0.010(0.084)$ & ${}^{+0.089}_{-0.069}$ \\
	&$\alpha$ & $-$ & ${-}^{}_{}$ & & $0.198(0.520)$ & ${}^{+0.333}_{-0.587}$ & & $-$ & ${-}^{}_{}$ & & $-$ & ${-}^{}_{}$ \\
	&$\beta$ & $-$ & ${-}^{}_{}$ & & $-0.097(0.629)$ & ${}^{+0.626}_{-0.625}$ & & $-$ & ${-}^{}_{}$ & & $-$ & ${-}^{}_{}$ \\
	\enddata
	\tablecomments{The marginalized mean values, standard deviations, and the 68\% CL of the flat $\Lambda$CDM model parameter and coefficients of $y_\mathrm{Amati}$, $y_\mathrm{exAmati}$, $y_\mathrm{copula_1}$ and $y_\mathrm{copula_2}$ from the A220 and A118 data set using the simultaneous fitting method. 
	}
\end{deluxetable}

\section{conclusions}\label{Sec_conclusion}
In this paper, we use the three dimensional Gaussian copula method  to investigate the luminosity correlation of GRB data. 
By assuming that the logarithms of  the special peek energy and the isotropic energy of GRBs  satisfy the Gaussian distributions and two different redshift distributions of GRB data (one is the special form given in Eq.~(\ref{power_z}) and the other is empirical distribution),  we obtain two improved Amati correlations of GRB data ($y_\mathrm{copula_1}$ and $y_\mathrm{copula_2}$), which are distinctively different from the standard Amati correlation and the extended Amati correlation.  After calibrating, with the low-redshift GRB data points from A220 and A118 data sets respectively,  these improved Amati correlations based on a fiducial $\Lambda$CDM model with $\Omega_\mathrm{m0}=0.3$ and $H_0=70 ~\mathrm{km~s^{-1}Mpc^{-1}}$,  and extrapolating the results to the  high-redshift GRB data,  we obtain the Hubble diagrams of 220 and 118 GRB data points. Applying these GRB data to constrain the $\Lambda$CDM model, we find that  the results from the improved Amati correlations  are apparently better  than those from  the standard Amati and extended Amati correlations although the BIC  favors mildly the standard Amati correlation. The improved Amati correlation based on the special redshift distribution of GRB data  gives the best  result, which is always consistent with  $\Omega_\mathrm{m0}=0.3$ at the $1\sigma$ confidence level and is highly consistent with  $\Omega_\mathrm{m0}=0.3$ when A220 data set is used. These results seem to indicate that when the improved Amati correlation with the special redshift distribution ($y_\mathrm{copula_1}$) is used in the low-redshift calibration, the GRB data can be regarded as a viable cosmological explorer.  However, the  BIC indicates that the standard Amati correlation remains to be favored mildly since it has the least model parameters. Furthermore,  once the simultaneous fitting method rather than the low-redshift calibration one is used,  we find that the constraints on $\Omega_{\mathrm {m0}}$ are weak and only the low bound limit on $\Omega_{\mathrm {m0}}$ can be obtained. Although this low bound limit from the improved Amati correlation is smaller than the one from the standard Amati correlation, there is no apparent evidence that the former is better than the latter. This result is different apparently from the one from the low-redshift calibration method.   Therefore, more works need to be done in the future in order to compare different Amati correlations.

\acknowledgments
We appreciate very much the insightful comments and helpful suggestions by anonymous referees.
This work was supported in part by the NSFC under Grants No. 12075084, No. 11690034, No. 11805063, No. 11775077, and 12073069,   
 by the Science and Technology Innovation Plan of Hunan province under Grant No. 2017XK2019, and by the Guizhou Provincial  Science and Technology Foundation (QKHJC-ZK[2021] Key 020).

\appendix
\section{The Amati correlation}\label{Amati}

In 2002,~\citeauthor{Amati2002} found that in GRB observational data there is a positive correlation between the spectral peak energy  $E_p$ and the isotropic equivalent radiated energy $E_{iso}$~\citep{Amati2006a,Amati2006b,Amati2008,Amati2009}, and this correlation has the form
\begin{equation}\label{2_Amati}
y_\mathrm{Amati}=a+bx,
\end{equation}
where
\begin{eqnarray}\label{epeisoxy}
y\equiv \log\frac{E_{iso}}{1\mathrm{erg}},~x\equiv \log\frac{E_p}{300\mathrm{keV}},
\end{eqnarray}
  intercept $a$ and slope $b$ are free coefficients, ``$\log$'' denotes the logarithm to base 10, and
\begin{eqnarray}\label{Ep}
E_p&=&E_p^{obs}(1+z),\\  \label{Eiso}
E_{iso}&=&4\pi d^2_L(z)S_{bolo}(1+z)^{-1}.
\end{eqnarray}
Here $z$ is the redshift, $E_P^{obs}$ is the observed peak energy of GRB spectrum, $d_L(z)$ is the luminosity distance, and $S_{bolo}$ is the bolometric fluence. 

If the coefficients $a$ and $b$ are determined, the luminosity distance of GRB data point can be obtained from Eqs.~(\ref{2_Amati}, \ref{Ep}, \ref{Eiso}). Then  we can obtain the distance modulus of GRB data, which is  defined to be
\begin{eqnarray}\label{mu}
\mu&=&5\log\frac{d_L(z)}{\mathrm{Mpc}}+25.
\end{eqnarray}

If assuming a fiducial cosmological model,   the values of coefficients $a$ and $b$ in the Amati correlation can be obtained from the observational data by using the following common fitting strategy~\citep{D'Agostini2005} :
\begin{eqnarray}\label{L}
\mathcal{L}(\sigma_{int},a,b)\propto\prod_{i}\frac{1}{\sqrt{\sigma_{int}^2+\sigma_{yi}^2+b^2\sigma_{xi}^2}}
\times\exp\left[-\frac{(y_i-a-bx_i)^2}{2(\sigma_{int}^2+\sigma_{yi}^2+b^2\sigma_{xi}^2)}\right],
\end{eqnarray}
where $\sigma_x$ and $\sigma_y$ are  the uncertainties of $x$ and $y$, respectively, and  $\sigma_{int}$ is the intrinsic uncertainty of GRB.
From  the well-known error propagation equation, one find that $\sigma _y$ and $\sigma_x$  can be derived from Eqs.~(\ref{Ep},\ref{Eiso}) and have the expressions:
\begin{eqnarray}\label{errepeiso}
\sigma_y=\frac{1}{\ln10}\frac{\sigma_{E_{iso}}}{E_{iso}},~
\sigma_x=\frac{1}{\ln10}\frac{\sigma_{E_{p}}}{E_{p}}
\end{eqnarray}
with
\begin{eqnarray}
\sigma_{E_{iso}}=4\pi d_L^2\sigma_{S_{bolo}}(1+z)^{-1}.
\end{eqnarray}
Here $\sigma_{E_{p}}$ and $\sigma_{S_{bolo}}$ are available in observations of GRBs.
Therefore,
maximizing the likelihood function $\mathcal{L}$ (Eq.~\ref{L}),  the allowed values of $a$, $b$, and $\sigma_{int}$ can be obtained.
Then the covariance matrix $C_{ij}$ of these fitted parameters can be approximately evaluated from:
\begin{equation}\label{covcal}
(C^{-1})_{ij}(\bm\theta_A)=\frac{\partial^2[-\ln\mathcal{L}(\bm{\theta}_A)]}{\partial\theta_i\partial\theta_j}\arrowvert_{\bm{\theta}_A=\bm{\theta}_b},
\end{equation}
where $\bm{\theta}_A=\left\{\sigma_{int},a,b\right\}$, and $\bm{\theta}_b$ denote the best-fitted value of $a$, $b$ and $\sigma_{int}$.

Using the best fitted values of $a$ and $b$, we can get the luminosity distance of GRBs and the corresponding distance modulus from Eq.~(\ref{mu}). By using the error propagation equation, the uncertainty of distance modulus can be derived from the following equation
\begin{eqnarray}\label{mu_err}
    \sigma_\mu^2=\left(\frac{5}{2}\sigma_{\log \frac{E_{iso}}{1\mathrm{erg}}}\right)^2+\left(\frac{5}{2\ln10}\frac{\sigma_{S_{bolo}}}{S_{bolo}}\right)^2.
\end{eqnarray}
Here
\begin{eqnarray}\label{errEiso}
\sigma_{\log \frac{E_{iso}}{1\mathrm{erg}}}^2&=&\sigma_a^2+\left(\sigma_b \log\frac{E_p}{300\mathrm{keV}}\right)^2+2\sum_{i=1}^{3}\sum_{j=i+1}^{3}\left(\frac{\partial y_\mathrm{Amati}(x;\bm{\theta}_A)}{\partial \theta_i}\frac{\partial y_\mathrm{Amati}(x;\bm{\theta}_A)}{\partial \theta_j}\right) C_{ij}\nonumber\\
&+&\left(\frac{b}{\ln 10}\frac{\sigma_{E_{p}}}{E_p}\right)^2+\sigma_{int}^2.
\end{eqnarray}

According to the distance modulus of GRBs, the cosmological model can be constrained by minimizing $\chi^2$
\begin{equation}\label{chi2}
    \chi^2=\sum_{i=1}^{N}\left[\frac{\mu_{obs}(z_i)-\mu_{th}(z_i; p)}
    {\sigma^{obs}_{\mu_i}}\right]^2,
\end{equation}
where $\mu_{obs}$ is the distance modulus of GRB and
$\mu_{th}(z_i;p)$ is the theoretic value of distance modulus in
cosmological  model with $p$ representing the model parameters. 

\section{The extended Amati correlation}\label{Appendix_exAmati}
The extended Amati correlation was proposed in \citep{Wang2017}  where the authors  used two formulas to parameterize the coefficients $a$ and $b$ in the standard Amati correlation:
\begin{eqnarray}\label{evoab}
a\rightarrow A=a+\alpha\frac{z}{1+z},~ b\rightarrow B=b+\beta\frac{z}{1+z},
\end{eqnarray}
where $\alpha$ and $\beta$ are two constants.
Substituting these parameterized formulas into the Eq.~(\ref{2_Amati}), the extended Amati correlation can be obtained
\begin{equation}\label{exAmati}
y_\mathrm{exAmati}=\left(a+\alpha\frac{z}{1+z}\right)+\left(b+\beta\frac{z}{1+z}\right)x.
\end{equation}
Recently, \cite{Khadka2021} used the A220 GRB data set to limit $\alpha$ and $\beta$, and found  that the Amati correlation is independent of redshift  within the error bars.  The coefficient $\bm{\theta}_{ex}=\{\sigma_{int},a,b,\alpha,\beta\}$ is also estimated from D'Agostinis likelihood function. The uncertainty of $\log\frac{E_{iso}}{\mathrm{1erg}}$ in Eq.~(\ref{mu_err}) can be obtained from 
\begin{eqnarray}
	\sigma_{\log \frac{E_{iso}}{1\mathrm{erg}}}^2&=&\sigma_a^2+\left(\sigma_b \log\frac{E_p}{300\mathrm{keV}}\right)^2
	+\left(\sigma_{\alpha}\frac{z}{1+z}\right)^2
	+\left(\sigma_{\beta}\frac{z}{1+z}\log\frac{E_p}{300\mathrm{keV}}\right)^2
	\nonumber\\
	&+&2\sum_{i=1}^{5}\sum_{j=i+1}^{5}\left(\frac{\partial y_\mathrm{exAmati}(x;\bm{\theta}_{ex})}{\partial \theta_i}\frac{\partial y_\mathrm{exAmati}(x;\bm{\theta}_{ex})}{\partial \theta_j}\right) C_{ij}\nonumber\\
	&+&\left[\left(b+\beta\frac{z}{1+z}\right)\frac{1}{\ln 10}\frac{\sigma_{E_{p}}}{E_p}\right]^2
	+\sigma_{int}^2.
\end{eqnarray}


\begin{thebibliography}{dummy}
	
	
	\bibitem[Abbott et al.(2018)]{Abbott2018} Abbott, T. M. C., et al. \href{https://doi.org/10.1093/mnras/sty1939}{2018, \mnras, 480, 3879}
	\bibitem[Aghanim et al.(2019)]{Planck} Aghanim, N., et al. \href{https://doi.org/doi:10.1051/0004-6361/201833880}{ 2020, \aap, 641, A1}
	\bibitem[Akaike (1974)]{Akaike1974} Akaike, H. \href{https://doi.org/doi:10.1109/TAC.1974.1100705}{ 1974, IEEE Trans. Autom. Control., 19, 716}
	\bibitem[Akaike (1981)]{Akaike1981} Akaike, H. \href{https://doi.org/10.1016/0304-4076(81)90071-3} {1981, J. Econom., 16,3}
	\bibitem[Amati et al.(2002)]{Amati2002} Amati, L., et al. \href{https://doi.org/10.1051/0004-6361:20020722}{2002, \aap, 390, 81}
	\bibitem[Amati (2006a)]{Amati2006a} Amati, L. \href{https://doi.org/10.1393/ncb/i2007-10064-9}{2006a, Nuovo Cimento B, 121, 1081}
	\bibitem[Amati (2006b)]{Amati2006b} Amati, L. \href{https://doi.org/10.1111/j.1365-2966.2006.10840.x}{2006b, \mnras, 372, 233}
	\bibitem[Amati et al.(2008)]{Amati2008} Amati, L., et al. \href{https://doi.org/10.1111/j.1365-2966.2008.13943.x}{2008, \mnras, 391, 577}
	\bibitem[Amati et al.(2019)]{Amati2019} Amati, L., D'Agostino, R., Luongo, O., Muccino, M., \& Tantalo, M. \href{https://doi.org/10.1093/mnrasl/slz056}{2019, \mnras, 486, L46}
	\bibitem[Amati et al.(2009)]{Amati2009} Amati, L., Frontera, F., \& Guidorzi, C. \href{https://doi.org/10.1051/0004-6361/200912788}{2009, \aap, 508, 173}
	
	
	
	
	\bibitem[Basilakos \& Perivolaropoulos (2008)]{Basilakos2008} Basilakos, S., \& Perivolaropoulos, L., \href{https://doi.org/10.1111/j.1365-2966.2008.13894.x}{2008, MNRAS, 391, 411}
	\bibitem[Benabed et al.(2009)]{Benabed2009} Benabed, K., Cardoso, J.-F., Prunet, S., \& Hivon, E. \href{https://doi.org/10.1111/j.1365-2966.2009.15202.x}{2009, \mnras, 400, 219}
	\bibitem[Birrer et al.(2020)]{Birrer2020} Birrer, S., et al. \href{https://doi.org/10.1051/0004-6361/202038861}{2020, \aap, 643, A165}
	
	\bibitem[Cao et al.(2022)]{Cao2022} Cao, S., Khadka, N., \& Ratra, B. \href{https://doi.org/10.1093/mnras/stab3559}{2022, \mnras, 510, 2928}
	\bibitem[Cao et al.(2021a)]{Cao2021a} Cao, S., Ryan, J., \& Ratra, B. \href{https://doi.org/10.1093/mnras/stab942}{2021a, \mnras, 504, 300}
	\bibitem[Cao et al.(2021b)]{Cao2021b} Cao, S., Ryan, J., Khadka, N., \& Ratra, B. \href{https://doi.org/10.1093/mnras/staa3748}{2021b, \mnras, 501, 1520}
	\bibitem[Chen et al.(2017)]{Chen2017} Chen, Y., Kumar, S., \& Ratra, B. \href{https://doi.org/10.3847/1538-4357/835/1/86}{2017, \apj, 835, 86}
	
	\bibitem[D'Agostini (2005)]{D'Agostini2005} D'Agostini, G. \href{https://arxiv.org/abs/physics/0511182v1}{2005,     arXiv:physics/0511182}
	\bibitem[Dekking et al.(2005)]{Dekking2005} Dekking, F. M., et al., \emph{A Modern Introduction to Probability and Statistics}. London, U.K.: Springer-Verlag, 2005
	\bibitem[Demianski et al.(2017)]{Demianski2017} Demianski, M., Piedipalumbo, E., Sawant, D., \& Amati, L. \href{https://doi.org/10.1051/0004-6361/201628909}{2017, \aap, 598, A112}
	\bibitem[Demianski et al.(2021)]{Demianski2021} Demianski, M., Piedipalumbo, E., Sawant, D., \& Amati, L. \href{https://doi.org/10.1093/mnras/stab1669}{2021, \mnras, 506, 903}
	\bibitem[Dirirsa et al.(2019)]{Dirirsa2019} Dirirsa, F. F, et al. \href{https://doi.org/10.3847/1538-4357/ab4e11}{2019, \apj, 887, 13}
	
	\bibitem[Efstathiou (2020)]{Efstathiou2020} Efstathiou, G. \href{https://arxiv.org/abs/2007.10716}{2020, arXiv:2007.10716v2}
	\bibitem[Eisenstein et al.(2005)]{BAO} Eisenstein, D. J., et al. \href{https://doi.org/doi:10.1086/466512}{2005, \apj, 633, 560.}
	
	
	
	\bibitem[Fenimore \& Ramirez-Ruiz (2000)]{Fenimore2000} Fenimore, E. E., \& Ramirez-Ruiz, E. \href{https://arxiv.org/abs/astro-ph/0004176}{2000, arXiv:astro-ph/0004176}
	
	\bibitem[Freedman (2021)]{Freedman2021} Freedman, W. L. \href{https://doi.org/10.3847/1538-4357/ac0e95}{2021, \apj, 919, 16}
	
	
	\bibitem[Ghirlanda et al.(2004a)]{Ghirlanda2004} Ghirlanda, G., Ghisellini, G., \& Lazzati, D. \href{https://doi.org/10.1086/424913}{2004a, \apj, 616, 331}
	\bibitem[Ghirlanda et al.(2004b)]{Ghirlanda2004b} Ghirlanda, G., Ghisellini, G., Lazzati, D., \& Firmani, C. \href{https://doi.org/10.1086/424915}{2004b, \apj, 613, L13}
	\bibitem[Ghirlanda et al.(2006)]{Ghirlanda2006} Ghirlanda, G., Ghisellini, G.,\& Firmani, C. \href{https://doi.org/10.1088/1367-2630/8/7/123}{2006, New,J. Phys., 8, 123}
	
	\bibitem[Hu et al.(2021)]{Hu2021} Hu, J. P., Wang, F. Y., \& Dai, Z. G. \href{https://doi.org/10.1093/mnras/stab2180}{2021, \mnras, 507, 730}
	
	
	
	
	
	
	\bibitem[Jiang et al.(2009)]{Jiang2009} Jiang,  I.-G., Yeh, L.-C., Chang, Y.-C., \& Hung, W.-L. \href{https://doi.org/10.1088/0004-6256/137/1/329}{2009, \aj, 137, 329}
	
	
	\bibitem[Khadka et al.(2021)]{Khadka2021} Khadka, N., Luongo, O., Muccino, M., \& Ratra, B. \href{https://doi.org/10.1088/1475-7516/2021/09/042}{2021, \jcap, 09, 042}
	
	\bibitem[Khadka \& Ratra (2020)]{Khadka2020} Khadka, N.,  \& Ratra, B. \href{https://doi.org/10.1093/mnras/staa2779}{2020, \mnras, 499, 391}
	
	\bibitem[Khetan et al.(2021)]{Khetan2021} Khetan, N., et al. \href{https://doi.org/10.1051/0004-6361/202039196}{2021, \aap, 647, A72}
	
	\bibitem[Klebesadel et al.(1973)]{Klebesadel1973} Klebesadel, R. W., Strong, I. B., \& Olson, R. A. \href{https://doi.org/10.1086/181225}{1973, \apj, 182, L85}
	
	\bibitem[Kodama et al.(2008)]{Kodama2008} Kodama, Y., et al. \href{https://doi.org/10.1111/j.1745-3933.2008.00508.x}{2008, \mnras, 391, L1}
	\bibitem[Koen (2009)]{Koen2009} Koen, C. \href{https://doi.org/10.1111/j.1365-2966.2008.14116.x}{2009, \mnras, 393, 1370}
	\bibitem[Koen \& Bere (2017)]{Koen2017} Koen, C., \& Bere,A. \href{https://doi.org/10.1093/mnras/stx1740}{2017, \mnras, 471, 2771}
	
	\bibitem[Li (2007)]{Li2007} Li, L.-X. \href{https://doi.org/10.1111/j.1745-3933.2007.00333.x}{2007, \mnras, 379, L55}
	\bibitem[Li et al. (2008)]{Li2008} Li, H. et al. \href{https://doi.org/10.1086/529582}{2008, \apj, 680, 92}
	\bibitem[Liang et al.(2008)]{Liang2008} Liang, N., Xiao, W. K., Liu, Y., \& Zhang, S. N. \href{https://doi.org/10.1086/590903}{2008, \apj, 685, 354}
	\bibitem[Liang et al.(2010)]{Liang2010} Liang, N., Wu, P.  \& Zhang, S. N. \href{https://doi.org/10.1103/PhysRevD.81.083518}{2010, \prd, 81, 083518}
	
	\bibitem[Lin et al.(2015)]{Lin2015}  Lin, H.-N., Li, X., Wang, S., \& Chang, Z. \href{https://doi.org/10.1093/mnras/stv1624}{2015, \mnras, 453, 128}
	\bibitem[Lin et al.(2016)]{Lin2016} Lin, H.-N., Li, X., Chang, Z. \href{https://doi.org/10.1093/mnras/stv2471}{2016, \mnras, 455, 2131}
	\bibitem[Lin \& Ishak (2021)]{Lin2021} Lin, W., \& Ishak, M. \href{https://doi.org/10.1088/1475-7516/2021/05/009}{2021, \jcap, 05, 009} 
	
	\bibitem[Luongo \& Muccino (2021)]{Luongo2021} Luongo, O., \& Muccino, M. \href{https://arxiv.org/abs/2110.14408}{2021, arXiv:2110.14408}
	
	
	\bibitem[Nelson (2006)]{Nelson2006} Nelson, R. B. 2006, An Introduction to Copulas (2nd ed.; New York: Springer)
	\bibitem[Norris et al.(2000)]{Norris2000} Norris, J. P., Marani, G. F., \& Bonnell, J. T. \href{https://doi.org/10.1086/308725}{2000, \apj, 534, 248}
	
	
	\bibitem[Perlmutter et al.(1999)]{Perlmutter1999} Perlmutter, S., et al. \href{https://doi.org/10.1086/307221}{ 1999, \apj, 517, 565}
	
	
	\bibitem[Qin et al.(2020)]{Qin2020} Qin, J., Yu, Y., \& Zhang, P. \href{https://doi.org/10.3847/1538-4357/ab952f}{2020, \apj, 897, 105}
	
	
	\bibitem[Riess et al.(1998)]{Riess1998} Riess, A. G., et al. \href{https://doi.org/10.1086/300499}{ 1998,  \aj, 116, 1009}
	\bibitem[Riess et al.(2018a)]{Riess2018} Riess, A. G., et al. \href{https://doi.org/10.3847/1538-4357/aaa5a9}{ 2018a, \apj, 853, 126}
	\bibitem[Riess et al.(2018b)]{Riess2018M} Riess, A. G., et al. \href{https://doi.org/doi:10.3847/1538-4357/aac82e}{ 2018b, \apj, 861, 126}
	\bibitem[Riess et al.(2021)]{Riess2021} Riess, A. G., et al. \href{https://doi.org/10.3847/2041-8213/abdbaf}{2021, \apjl, 908, L6}
	
	
	
	\bibitem[Sato et al.(2010)]{Sato2010} Sato, M., Ichiki, K., \& Takeuchi, T. T. \href{https://doi.org/10.1103/PhysRevLett.105.251301}{2010, \prl, 105, 251301}
	\bibitem[Sato et al.(2011)]{Sato2011} Sato, M., Ichiki, K., \& Takeuchi, T. T. \href{https://doi.org/10.1103/PhysRevD.83.023501}{2011, \prd, 83, 023501}
	\bibitem[Scherrer et al.(2010)]{Scherrer2010} Scherrer, R. J., Berlind, A. A., Mao, Q., \& McBride, C. K. \href{https://doi.org/10.1088/2041-8205/708/1/L9}{2010, \apjl, 708, L9}
	\bibitem[Schwarz (1978)]{Schwarz1978} Schwarz, G. \href{https://www.jstor.org/stable/2958889}{1978, Annals of Stats, 6, 461}
	\bibitem[Spergel et al.(2003)]{CMB1} Spergel, D. N., et al. \href{https://doi.org/doi:10.1086/377226}{ 2003, \apjs, 148, 175}
	\bibitem[Spergel et al.(2007)]{CMB2} Spergel, D. N., et al. \href{https://doi.org/10.1086/513700}{ 2007, \apjs, 170, 377}
	
	
	\bibitem[Takeuchi (2010)]{Takeuchi2010} Takeuchi, T. T. \href{https://doi.org/10.1111/j.1365-2966.2010.16778.x}{2010, \mnras, 406, 1830}
	\bibitem[Takeuchi \& Kono (2020)]{Takeuchi2020} Takeuchi, T. T., \& Kono, K. T. \href{https://doi.org/10.1093/mnras/staa2558}{2020, \mnras, 498, 4365}
	\bibitem[Takeuchi et al.(2013)]{Takeuchi2013} Takeuchi, T. T., Sakurai, A., Yuan, F.-T., Buat, V., \& Burgarella, D. \href{https://doi.org/10.5047/eps.2012.06.008}{2013, Earth Planet Sp, 65, 281}
	
	
	\bibitem[Wang et al.(2015)]{Wang2015} Wang, F. Y., Dai, Z. G., \& Liang, E. W. \href{https://doi.org/10.1016/j.newar.2015.03.001}{2015, New Astronomy Reviews, 67, 1}
	\bibitem[Wang et al.(2021)]{Wang2021}  Wang, F. Y., Hu, J. P., Zhang, G. Q., \& Dai, Z. G. \href{https://doi.org/10.3847/1538-4357/ac3755}{2022, \apj, 924, 97}
	\bibitem[Wang et al.(2011)]{Wang2011} Wang, F.-Y., Qi, S., \& Dai, Z.-G. \href{https://doi.org/10.1111/j.1365-2966.2011.18961.x}{2011, \mnras, 415, 3423}
	\bibitem[Wang et al.(2017)]{Wang2017} Wang, G.-J., Yu, H., Li, Z.-X., Xia, J.-Q., \& Zhu Z.-H. \href{https://doi.org/10.3847/1538-4357/aa5b9b}{2017, \apj, 836, 103}
	
	\bibitem[Wang et al.(2016)]{Wang2016}  Wang, J. S., Wang, F. Y., Cheng, K. S. \& Dai, Z. G. \href{https://doi.org/10.1051/0004-6361/201526485}{2016, \aap, 585, A68}
	\bibitem[Wei \& Zhang.(2009)]{Wei2009} Wei, H., Zhang, S. N. \href{https://doi.org/10.1140/epjc/s10052-009-1086-z}{2009, Eur. Phys. J. C, 63, 139}
	\bibitem[Wu et al.(2017)]{Wu}Wu, P.-X, Li, Z.-X, \& Yu, H.-W. \href{https://doi.org/10.1007/s11467-016-0599-9}{2017, Front. Phys. 12, 129801}
	
	\bibitem[Yonetoku et al.(2004)]{Yonetoku2004} Yonetoku, D., et al. \href{https://doi.org/10.1086/421285}{2004, \apj, 609, 935}
	\bibitem[Yuan et al.(2018)]{Yuan2018} Yuan, Z., Wang, J., Worrall, D. M., Zhang, B.-B., \& Mao, J. \href{https://doi.org/10.3847/1538-4365/aaed3b}{2018, \apjs, 239, 33}
	
	
\end{thebibliography}
\end{document}